# Atomic Insights into the Oxidative Degradation Mechanisms of Sulfide Solid Electrolytes

**Authors:**

Chuntian Cao[1,†], Matthew R. Carbone[1,†], Cem Komurcuoglu[2], Jagriti S. Shekhawat[2], Kerry Sun[2], Haoyue Guo[2], Sizhan Liu[3], Ke Chen[3], Seong-Min Bak[4,5], Yonghua Du[4], Conan Weiland[6], Xiao Tong[7], Dan Steingart[2,8,9], Shinjae Yoo[1], Nongnuch Artrith[10], Alexander Urban[2,8,*], Deyu Lu[7,*], Feng Wang[3,*,a]

[1] Computational Science Initiative, Brookhaven National Laboratory, Upton, NY 11973, USA

[2] Department of Chemical Engineering, Columbia University, New York, NY 10027, USA

[3] Interdisciplinary Science Department, Brookhaven National Laboratory, Upton, NY 11973, USA

[4] National Synchrotron Light Source II, Brookhaven National Laboratory, Upton, NY 11973, USA

[5] Department of Materials Science and Engineering, Yonsei University, Seoul 03722, Republic of Korea

[6] Material Measurement Laboratory, National Institute of Standards and Technology, Gaithersburg, MD 20899, USA

[7] Center for Functional Nanomaterials, Brookhaven National Laboratory, Upton, NY 11973, USA

[8] Columbia Electrochemical Energy Center (CEEC), Columbia University, New York, NY 10027, USA

[9] Department of Earth and Environmental Engineering, Columbia University, New York, NY 10027, USA

[10] Materials Chemistry and Catalysis, Debye Institute for Nanomaterials Science, Utrecht University, 3584 CG Utrecht, The Netherlands

Equal contribution: †

[a] Present address: Applied Materials Division, Argonne National Laboratory, 9700 S. Cass Avenue, Lemont, IL 60439

**Abstract:**

Electrochemical degradation of solid electrolytes is a major roadblock in the development of solid-state batteries, and the formed solid-solid interphase (SSI) plays a key role in the performance of solid-state batteries. In this study, by combining experimental X-ray absorption spectroscopy (XAS) measurements, first-principles simulations, and unsupervised machine learning, we have unraveled the atomic-scale oxidative degradation mechanisms of sulfide electrolytes at the interface using the baseline $Li_3PS_4$ (LPS) electrolyte as a model system. The degradation begins with a decrease of Li neighbor affinity to S atoms upon initial delithiation, followed by the formation of S-S bonds as the $PS_4$ tetrahedron deforms. After the first delithiation cycle, the $PS_4$ motifs become strongly distorted and $PS_3$ motifs start to form. Spectral fingerprints of the local structural evolution are identified, which correspond to the main peak broadening and the peak shifting to a higher energy by about 2.5 eV in P K-edge XAS and a new peak emerging at 2473 eV in S K-edge XAS during delithiation. The spectral fingerprints serve as a proxy for the electrochemical stability of phosphorus sulfide solid electrolytes beyond LPS, as demonstrated in argyrodite $Li_6PS_5Cl$. We observed that the strong distortion and destruction of $PS_4$ tetrahedra and the formation of S-S bonds are correlated with an increased interfacial impedance. To the best of our knowledge, this study showcases the first atomic-scale insights into the oxidative degradation mechanism of the LPS electrolyte, which can provide guidance for controlling macroscopic reactions through microstructural engineering and, more generally, can advance the rational design of sulfide electrolytes.

**Highlights:**

- Data-driven X-ray absorption spectroscopy (XAS) analysis combines structural and spectral databases and unsupervised machine learning
- XAS fingerprints identify local structure evolution upon electrochemical degradation
- $Li_3PS_4$ oxidation leads to distortion and breakdown of $PS_4$ tetrahedra and formation of S-S bonds
- Microscopic structural changes suggest the possible origin of observed macroscopic impedance increase

**Introduction:**

Sulfide-based solid electrolytes (SEs) are promising candidates for solid-state batteries because of their high ionic conductivity and malleability[1-3]. In conventional liquid-electrolyte batteries, a solid electrolyte interphase layer forms at the electrolyte/electrode boundaries due to electrolyte decomposition outside its electrochemical stability window and interface reactions with the electrode surface[4]. Similarly, SEs can also undergo (de)lithiation in contact with the positive (negative) electrode active materials to form solid-solid interphase (SSI) layers between the SE and the electrodes[3, 5, 6]. Compared to oxide-based SEs, sulfide-based SEs have a significantly narrower electrochemical stability window[7] and suffer from electrochemical decomposition when the potential is outside this range[8]. Additionally, the interfaces between sulfide SEs and several anode/cathode active materials are believed to be chemically unstable[9-11]. It is essential to elucidate the mechanism of the interfacial chemical and electrochemical reactions and the atomic details of the formed SSI layer to improve the performance of solid-state batteries.

The ideal SSI layer is ionically conductive and electronically insulating so that it can passivate the interface and prevent further electrolyte decomposition while still allowing $Li^+$ transport. However, this is often not the case in sulfide SEs. In terms of ionic conductivity, a previous first-principles study predicted interface reaction products under different voltage[9], and suggest they may increase the interfacial resistance and negatively impact the cell performance[12, 13]. Additionally, SSIs may be poor electronic insulators that do not completely passivate the interface, leading to a continued growth of the SSI, the loss of cyclable $Li^+$, and poor Coulombic efficiency[14]. An X-ray photoelectron spectroscopy (XPS) study on Li anodes[15] and an electrochemical impedance spectroscopy (EIS) study on $LiCoO_2$ cathodes[16] have shown that the interphase layer continues to grow throughout the charging process, and the charge transfer resistance increases during cycling. In addition, the decomposition products may be electrochemically active and participate in battery cycling[17, 18], which adds to the complexity of analyzing the charge/discharge process. Understanding the formation and growth of the SSI is crucial to tuning its properties as SEs are a promising candidate in future commercial solid-state batteries.

The growth, composition, and properties of the SSI layer have drawn great attention. Experimental methods like electrochemical characterization[8], XPS[15, 19-26], X-ray diffraction (XRD)[27], X-ray absorption spectroscopy (XAS)[28-31], Raman spectroscopy[32-34], and cryogenic electron microscopy (cryo-EM)[35] have been used to characterize the SSI layer. First-principles theories are widely used to calculate the electrochemical stability windows of SEs[7, 11, 36, 37]. It has been found that sulfide SEs decompose upon electrochemical delithiation and form oxidized sulfur species[8]. *In situ* XPS studies have investigated the interphase growth of sulfide SEs in contact with Li metal[21, 22] and observed the formation of resistive interfaces composed of $Li_2S$ and $Li_3P$. A previous XAS study[28] indicated that bridging S-S bonds between $PS_4$ units in sulfide electrolytes associate and dissociate during the delithiation and lithiation processes, respectively, but further atomic-scale mechanistic understanding is lacking. Beyond the characterization studies, engineering approaches aimed at improving the interface properties by adding a buffer layer between the electrolyte and the electrode[16, 38-40], or changing the composition of the sulfide electrolyte to tune the SSI properties[41-44], have been widely explored.

Despite various experimental studies of the behavior of the interphase layer and the computational prediction of thermodynamic decomposition products, there still lacks a fundamental understanding of the kinetically driven interface reaction mechanism on the atomic scale, which is essential to controlling the interface reactions that promote the formation of ideal interphase layers. This knowledge can help explain the redox reactions of sulfide electrolytes and guide the design of electrolytes and buffer layers. Here, we, therefore, interrogate the standard sulfide electrolytes to obtain a thorough mechanistic understanding of the interface reactions.

In this study, we combine experimental XAS measurements, first-principles structural modeling and spectral simulations, and unsupervised machine learning (ML) to investigate the atomic-scale oxidative degradation process of a prototypical sulfide SE. XAS has previously been used to study the structural and electronic state changes of $Li_3PS_4$ during lithiation and delithiation[28], and to study the structural stability of $Li_{10}GeP_2S_{12}$ at high voltages[30]. But interpreting the X-ray absorption near edge structure (XANES) often requires extensive spectral modeling on a variety of relevant atomic structures[45, 46]. As a result, it is highly non-trivial to disentangle XANES spectra and gain fundamental insights on local structural motifs. Recently, ML methods have been successfully applied to XANES analysis for extracting local descriptors (e.g. average coordination number of metal nanoparticles[47] and the local chemical environment of 3d transition-metal elements[48, 49]) from measured XANES spectra and for high-throughput prediction of molecular XANES spectra from atomic structures[50]. In this study, we used first-principles simulations to obtain XANES fingerprints and employed unsupervised ML to correlate these spectral features with local structural motifs. This enables the development of a coherent picture of the interface reaction.

Specifically, we focus on the baseline $Li_3PS_4$ (LPS) electrolyte to study its structural evolution upon delithiation and the formation of SSI layer. Experimentally, we measured the S and P K-edge XAS spectra at different delithiation states in different cycles. To analyze the experimental spectra, we developed an atomic structural database using density functional theory (DFT), which contains 660 configurations with varying Li content. We then simulated the XANES spectra for all the symmetrically inequivalent P (2227 in total) and S sites (8885 in total) in the database with first-principles methods. Using the structural and spectral databases, we applied unsupervised ML to correlate the spectral features with structural motifs. By combining first-principles simulation and ML, established trends from the simulation databases can successfully interpret the experimental data and yield a physics-based model to unravel the intriguing structure-property relationship of the LPS SSI.

We have found that at the microscopic scale, the delithiation starts in the first cycle with a loss of Li neighbors around S atoms and the subsequent formation of S-S bonds along with the distortion of the $PS_4$ tetrahedra. In the second and following cycles, the $PS_4$ motifs are strongly distorted and $PS_3$ motifs form. The strong $PS_4$ distortions, the formation of S-S bonds and the breakage of the $PS_4$ tetrahedra correspond to distinct P and S K-edge XANES fingerprints in the experiment. The destruction of the $PS_4$ motifs and the formation $PS_3$ motifs provide the likely microscopic origin of the increased interfacial impedance during battery cycling. To the best of our knowledge, this study is the first to shed light on the atomic-scale mechanism of LPS oxidative degradation by correlating the XANES spectral fingerprints with structural and chemical descriptors.

As a further step, we demonstrated the significance of the identified spectral fingerprints in XANES analysis of other phosphorus sulfide electrolytes. We found that the distortion of the $PS_4$ tetrahedra in argyrodite $Li_6PS_5Cl$ electrolyte is greatly alleviated compared to LPS, which could explain the improved interfacial stability of argyrodite electrolytes[51, 52]. This suggests that the spectral fingerprints also serve as a proxy of the electrochemical stability of phosphorus sulfide solid electrolytes. Our results can provide important guidance to the development of sulfide electrolytes and controlling macroscopic interface reactions through microstructural engineering in electrolyte design. Moreover, the combined simulation and ML-assisted XAS analysis approach developed in this work can be generalized to study a broad range of energy storage materials with high potential impact on XAS-based materials characterization.

## Results and Discussion:

### Experiment Setup

We measured the electrochemical redox reactions and X-ray absorption spectra on LPS-C composite, which consists of LPS particles and carbon black at a ratio of 7:3 by mass. The electrochemistry experiments were carried out in an LPS-C | LPS | Li-In cell as shown in Figure 1 (a). The LPS-C composite is the working electrode and lithium-indium alloy is the counter/reference electrode. More details about the fabrication of the LPS-C composite and cell assembly can be found in the Methods section.

The use of LPS-C composite increases the surface-to-volume ratio because the redox reactions occur at the surface of every LPS particle in contact with carbon. This helps to improve the signal-to-noise ratio in XAS measurements. In general, solid state batteries use a solid electrolyte-cathode active material composite as the catholyte[18, 53]. Our LPS-C composite is similar to the catholyte except that the inactive carbon black is used instead of the cathode material, because we would like to study the intrinsic decomposition without interference from the chemical reactions between LPS and the active material.

### Electrochemistry

We performed cyclic voltammetry (CV) experiments in an inert atmosphere glovebox under the voltage sweep rate of 0.1 mV/s on the LPS-C | LPS | Li-In cell, as shown in Figure 1 (b). The electrochemistry starts with an anodic scan, i.e., an increase in the voltage from the open-circuit voltage (OCV) to 5 V vs. $Li^+$/Li. The OCV for this cell is at 1.65 V vs. $Li^+$/Li. We define one cycle as the voltage sweep OCV → 5 V → 0 V → OCV.

We first observe a large current bump labeled with (i), which starts at around 2.5 V and has a maximum current at around 3.6 V. Here, we use the term "bump" instead of "peak" because of the relatively large width as compared to common CV sweep peaks. This bump is indicative of an oxidative decomposition reaction, likely associated with the delithiation of LPS. During the reverse scan from 5 V to 0 V, the current decreases as the voltage decreases. At around 2.4 V, the current becomes negative, and its amplitude increases as the voltage decreases, corresponding to the lithiation reaction. In the sweep from 0 V to 5 V, the current turns positive at about 0.7 V. There are two positive current bumps at around 1.35 V (ii) and 3.25 V (iii), respectively, indicating two different delithiation reactions. We note that the bump at 3.25 V (iii) is at a different position from the 3.6 V bump (i). The delithiation reaction of the first and the second cycles are different, and the voltage curves after the second cycle are similar. This implies that the redox reactions in the first cycle are different from the following cycles, which may be caused by the composition change of the LPS after the first cycle. To further investigate the mechanism of the different redox reactions, we performed XAS experiments as well as a computational delithiation of LPS with first-principles calculations, as explained in following sections.

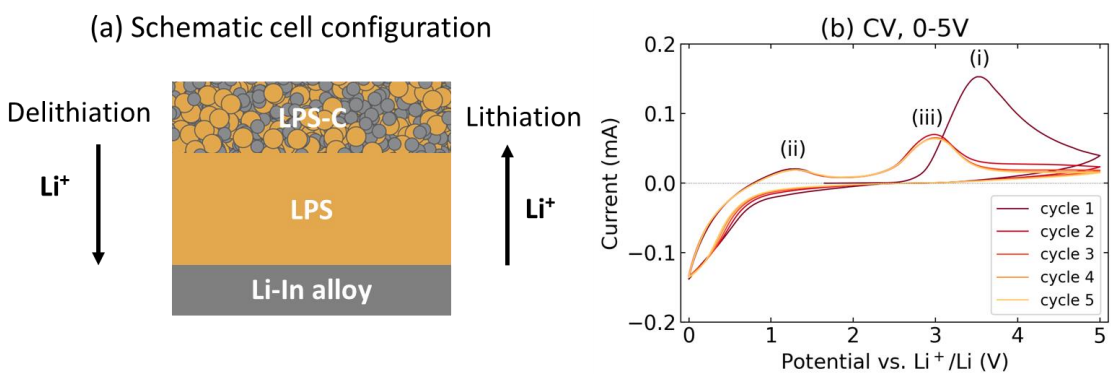

Figure 1. Cyclic voltammetry (CV) experiments of LPS-C | LPS | Li-In cells under the voltage scan rate of 0.1 mV/s. (a) Illustration of the all-solid-state battery using LPS-C working electrode, Li-In alloy counter/reference electrode, and LPS electrolyte pellet. (b) Electrochemistry profile of a CV experiment starting with an anodic sweep from the open-circuit voltage (OCV) to 5 V vs. Li$^+$/Li and then reverses the scan to 0 V. Accuracy of the current measurement is expected to be better than 1 μA and voltage better than 1 mV.

*Experimental X-ray absorption near edge structure spectroscopy (XANES)*

We measured the P and S K-edge X-ray absorption near edge structure spectroscopy (XANES) of a series of LPS-C composite electrodes cycled to different charge stages in the 0 V to 5 V range, as shown in Figure 2. At each charge state, two identical samples were prepared and measured using two detection modes, electron yield (EY) and fluorescence yield (FY), at Beamlines SST-2 and TES at the National Synchrotron Light Source II (NSLS-II), respectively. The detection depths of EY and FY are about 10 nm and 5 μm, and the particle size of LPS is 2 μm to 10 μm (Figure S1). Thus, EY and FY provide information on the surface and bulk of the sample, respectively.

The spectra measured in the first delithiation cycle from open circuit voltage (OCV) to 5 V include (0) pristine LPS, (1) cycle1: 3.4 V, and (2) cycle1: 5 V spectra, as shown in Figure 2. At 3.4 V, the delithiation has just started, and the change in the spectra is minimal. The only observable feature is a slight intensity increase at 2473 eV in the S spectrum, as indicated by the purple arrow in Figure 2 (b). At the end of the first delithiation (5 V), the P K-edge XANES white line peak at 2148.5 eV decreases and broadens. In the S

K-edge spectrum, a peak at 2473 eV appears as indicated by the blue arrows in Figure 2 (b) and (d), which can be attributed to the formation of bridging S bonds [8, 54].

Spectra (3) cycle 1: 0 V and (4) cycle 2: 2.3 V are from the first lithiation process. Although the change from (3) to (4) is from the anodic sweep 0 V to 2.3 V, (4) is still classified as the end of lithiation rather than the midpoint of the second delithiation because the net capacity from (3) to (4) is negative. The intensity of P XANES white line increases slightly but does not reach the original intensity in pristine LPS spectrum. The peak at 2473 eV in S spectra decreases, and the S4 spectra in both EY and FY show a three-peak feature similar to the pristine LPS spectrum, but with a higher intensity in the second peak region.

During the second delithiation from (4) to (5), a new peak at 2151 eV emerges in the P5 spectra, as indicated by the orange arrows in Figure 2 (a) and (c). This new spectral feature indicates different reaction mechanisms between the first two delithiation processes, which is also reflected in the different current peak positions in the CV experiment (Figure 1 (b)). The change in the S spectra is similar to the first delithiation. During the second lithiation, the changes in P and S are (partially) reversible, with both peaks at 2151 eV of P and 2473 eV of S decrease in P6 and S6.

The EY and FY experiments exhibit the same overall trend of spectral evolution with several subtle differences. The first difference is that the intensities of the white line peaks in EY spectra are higher. This is because the peaks in FY spectra are dampened by self-absorption while the EY spectra do not have self-absorption. The second difference comes from the different probe depths. Taking the S2 spectra as an example, the ratio between the new 2473 eV peak and the original white line peak is much higher in EY than in FY, indicating that the reaction occurs mainly at the surface of LPS particles. The signal from unreacted bulk LPS is negligible in EY, while FY measures the mixed signal from both bulk and surface. A similar behavior can be found in the S5 and P5 spectra. To further analyze the key spectral features in XANES, we combine first-principles calculations and machine learning to reveal the atomic scale structural details of the LPS redox reactions.

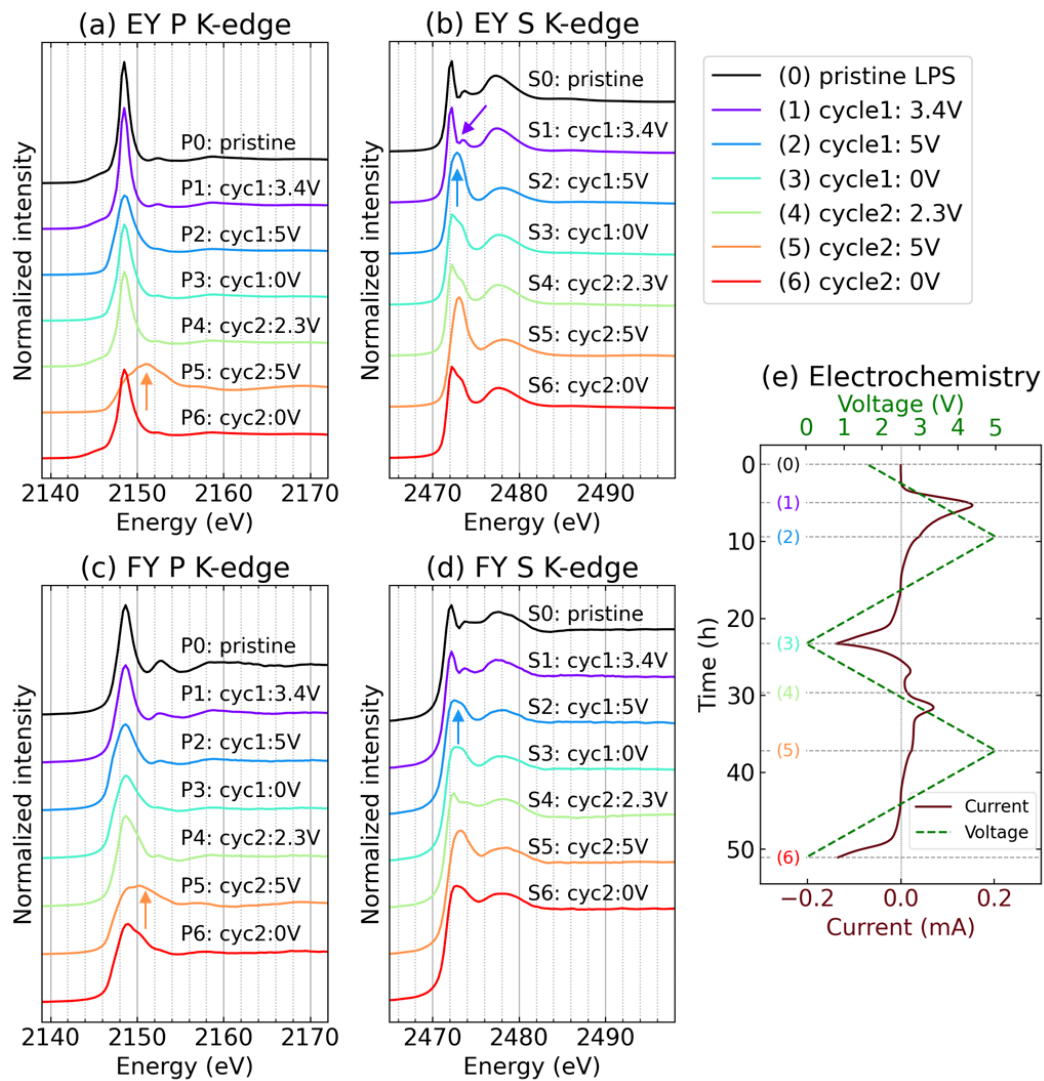

*Figure 2. Experimental X-ray absorption near edge structure (XANES) data collected on LPS-C composite electrodes. LPS-C electrodes were cycled to different states in 0 V to 5 V CV experiment and measured ex situ after cell disassembly. (a) P K-edge and (b) S K-edge spectra measured in electron yield (EY) mode at SST-2 beamline. (c) P K-edge and (d) S K-edge spectra measured in fluorescence yield (FY) mode at TES beamline. (e) Current (red solid line) and voltage (green dashed line) profiles of the CV experiment with 0.1 mV/s scan rate; the horizontal dashed lines indicate the states that the samples were prepared for XAS measurement. Error in the measured XANES intensity is better than 0.1% and in photon energy better than 0.05 eV. The error in the measured CV voltage is better than 1 mV and current less than 1 μA.*

### Computational delithiation of β-Li₃PS₄

To aid with the interpretation of the above experimental characterization of the electrochemical degradation of LPS, we performed a computational delithiation of β-$Li_3PS_4$ with DFT calculations. We first systematically enumerated distinct lithium/vacancy orderings in super cells with the general composition $Li_{12-x}P_4S_{16}$ (*i.e.*, four LPS formula units), and then repeated the delithiation a second time sequentially, such that in each step one more Li atom was removed from the relaxed structure, approximating a kinetically controlled lithium extraction. In total the process yielded 660 atomic configurations with varying lithium contents that were fully optimized (atomic positions and lattice parameters). Further details of the computational delithiation are given in the Methods section.

Figure 3 (a) shows the formation energies of the delithiated LPS structures. Three intermediate compositions are predicted to be thermodynamically stable at 0 Kelvin, $LiPS_4$, $Li_{0.5}PS_4$, and $Li_{0.25}PS_4$, though the convex hull is shallow and additional compositions might be accessible at room temperature. Based on the calculations, LPS is fully delithiated at potentials above approximately 3.4 V vs. Li/Li$^+$. In $\beta$-$Li_3PS_4$, sulfur and phosphorus atoms form isolated tetrahedral $PS_4$ motifs. Upon lithium extraction, we find that the $PS_4$ tetrahedra form chains via S-S bonding (Figure 3 (b) and (c)). The number of bonds per $PS_4$ unit increases with the degree of delithiation. Furthermore, with increasing degree of delithiation, S-S bonds can form within a $PS_4$ tetrahedron (Figure 3 (d)), and $PS_3$ structure motifs are observed in deeply delithiated stoichiometries, e.g., $Li_{0.25}PS_4$, as shown in Figure 3 (e).

The S-S bonding indicates that sulfur oxidizes when lithium is extracted from LPS. This mechanism is corroborated by a Bader charge analysis, suggesting that only sulfur atoms participate in the redox reaction and gradually oxidize from initially $S^{2-}$ in $Li_3PS_4$ to a 1:3 mixture of $S^{2-}$ and $S^-$ in the fully delithiated $P_4S_{16}$. The calculations do not show any evidence of phosphorus oxidation, which remains in 5+ valence state throughout.

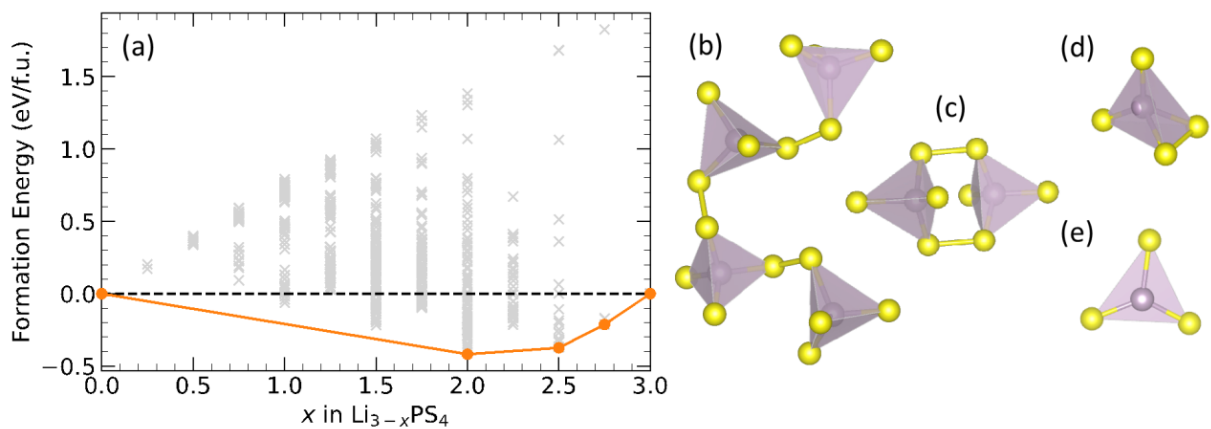

Figure 3. (a) Calculated formation energies of $Li_{12-x}P_4S_{16}$ structures with enumerated lithium/vacancy orderings. Those structures with formation energies on the lower convex hull (orange line) are predicted to be thermodynamically stable at zero Kelvin relative to the reference compositions. But given the shallow depth of the hull, other compositions might be observable at room temperature. (b – c) $PS_x$ structure motifs found in the stable delithiated structures (sulfur yellow, phosphorus violet). (d – e) $PS_4$ and $PS_3$ motifs in deeply delithiated structures.

### Unsupervised machine learning analysis of P K-edge and S K-edge XANES

We constructed simulated P and S K-edge XANES databases based on the delithiated β-LPS ($Li_{12-x}P_4S_{16}$) compounds described in the previous section. The spectra databases contain 2227 P and 8885 S spectra calculated from symmetrically unique sites. The spectral calculations were benchmarked on compounds with known experimental spectra standards confirming the predictive power of our approach[55].

In this step, we focus on distinguishing different *relative* spectral shapes. Thus, the simulated spectra are first aligned to the first major peak position and then clustered into different groups using the K-Means algorithm[56]. Edge alignment using a first-principles procedure is applied subsequently to the clustered spectra. This procedure isolates the numerical error in the spectral clustering and the residue systematic error in the edge alignment, which makes our analysis pipeline more robust. The spectra in each cluster after the edge-alignment procedure are plotted in the Supporting Information, Figures S6 and S7.

We used five groups in P clustering and six groups in S clustering. Details of the spectra simulation and machine learning can be found in the Methods section. The clustering results of P and S XANES spectra are shown in Figure 4 and Figure 6, respectively. The clusters are arranged in the descending order by Li stoichiometry.

*Clustering of P K-edge XANES*

Figure 4 shows the clustering result of simulated P K-edge XANES data. Spectra in Figure 4 (a) from Cluster 1 correspond to a nearly ideal tetrahedral $PS_4$ motif in β-LPS, consistent with the strong single peak in Figure 2 (a) and (c) of pristine β-LPS spectra P0. The trend in Figure 4 (b – e) shows the splitting of single peak into two, which, notwithstanding various broadening effects (e.g. the finite-temperature broadening), are consistent with the features in Figure 2 (a) and (c), spectra P5 at cycle 2, 5 V.

To relate the spectral changes to structural changes, we plot several key structural descriptors averaged within each cluster as shown in Figure 4 (g – j), including Li concentration, minimum and maximum S-P-S bond angles in each polyhedron, S-P bond length and the number of neighboring S atoms in the P polyhedral. We also performed complementary structural analysis using Smooth Overlap of Atomic Positions (SOAP) descriptors[57], and the results are summarized in Figure S4 in the Supporting Information. Figure 4 (g) shows that the overall Li count, both stoichiometrically and present in the second coordination shell of P (the first coordination shell of P only contains S atoms), decreases monotonically with respect to the cluster index. This confirms that the order of the cluster tracks the degree of the delithiation from low to high.

The trends of the structural descriptors are strongly correlated with the spectral trends going from Cluster 1 to 5. First, while the maximum S-P-S bond angle stays mostly constant, the minimum bond angle decreases, indicating a distortion of the $PS_4$ structure in which S atoms bonded to the central P atom are moving into closer proximity. On the spectra side, in Cluster 2 to 5 the single peak in the spectra splits into two peaks, which is correlated with the distortion of $PS_4$ tetrahedra in the structure descriptor. Second, the largest S-P bond length of the $PS_4$ motif increases from approximately 2.05 Å in Cluster 1 to 2.40 Å in Cluster 5 during the delithiation. In fact, some of the P sites in Cluster 5 only have three S neighbors, as indicated in Figure 4 (j), representing $PS_3$ motifs.

By correlating structure descriptors with spectral features, we can identify the pathway of the local structural evolution of P polyhedra upon delithiation. In Cluster 2 and 3, the minimum S-P-S bond angle decreases, indicating that two S atoms are moving closer to each other. The peak broadening from the simulation is similar to the experimental spectrum after the 1st delithiation. In Cluster 4, the minimum S-P-S bond angle decreases by almost a factor of 2, forming a sharp S-P-S angle of approximately 70 degrees, and the spectra also show a larger splitting. In Cluster 5, some $PS_4$ tetrahedra are severely distorted, forming $PS_3$ motifs. These strongly distorted P local motifs have a spectral fingerprint, characterized by a second peak at 2151 eV in P K-edge XANES after the 2nd delithiation. We conclude that the new peak comes from a mixture of distorted $PS_4$ motifs.

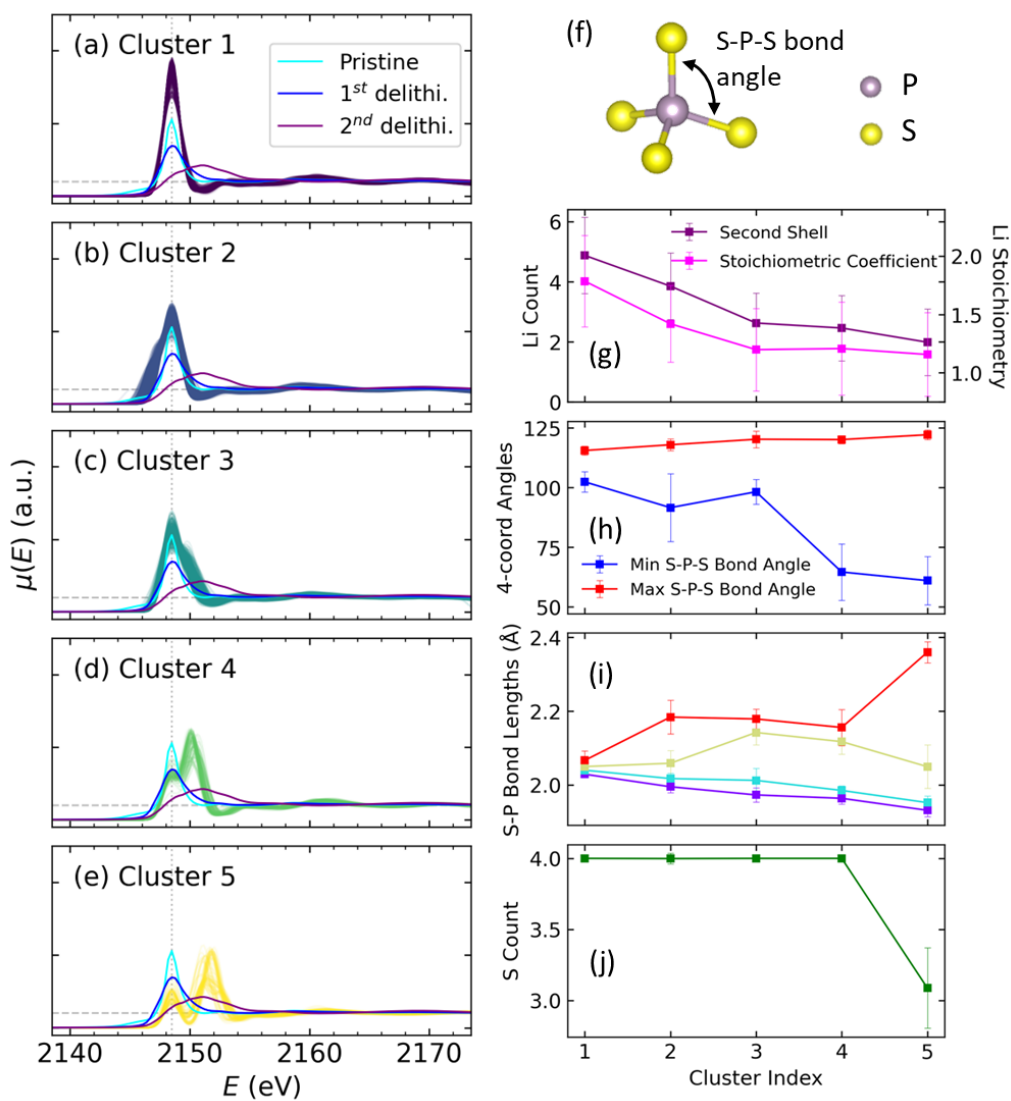

*Figure 4. (a – e) Simulated P K-edge XANES spectra of delithiated β-LPS clustered into five groups using K-Means. (f) Illustration of the $PS_4$ local structure motif. (g – j) Averaged structural descriptors in each of the 5 clusters. The markers show the mean values of the descriptors within each cluster with error bars representing the standard deviation. (g) Lithium stoichiometry and the number of Li atoms in the second shell, (h) minimum and maximum S-P-S bond angles, (i) S-P bond length, and (j) the number of neighboring S atoms.*

We compare the P K-edge XANES of representative simulated P site-spectra with experiment to understand the structural and electronic origin of the XANES spectral features in Figure 5. In particular, we choose three representative local motifs: 1) a nearly ideal $PS_4$ motif (distortion index[58, 59]: 0.007; bond angle variance: 21.88; Min S-P-S angle: 104.54°; Max S-P-S angle: 115.32°) from the pristine $Li_3PS_4$, 2) a largely distorted $PS_4$ motif (distortion index: 0.049; bond angle variance: 587.37; Min S-P-S angle: 57.98°; Max S-P-S angle: 121.04°) from a delithiated $LiPS_4$ structure, and 3) a $PS_3$ motif from a deeply delithiated $Li_{0.5}PS_4$ structure with the surface normal direction along $\hat{z}$, as shown in Figure 5 (b).

The simulated pristine Li$_3$PS$_4$ spectrum has a single pronounced main peak at 2148.5 eV, in good agreement with the spectral shape of experiment. Further projected density of states (PDOS) analysis of the P atom and its four neighboring S atoms shows that the main peak comes from the P-S antibonding σ* orbitals in the tetrahedral local symmetry. In the distorted PS$_4$, due to the breaking of the T$_d$ symmetry, the σ* band is broadened and becomes split as shown in the P and S 3*p* PDOS in Figure 5 (d). Accordingly, the spectrum of distorted PS$_4$ shows a qualitatively different spectral line shape from the pristine Li$_3$PS$_4$ with a peak splitting of 2 eV. Figure S9 plots the PDOS of slightly distorted PS$_4$ structures, showing a smaller peak splitting of about 1.5 eV. The PS$_3$ motif in deep delithiated LPS exhibits a double peak feature at around 2151 eV and another peak at lower energy around 2147.9 eV. The simulated spectrum of the PS$_3$ motif is in good agreement with the broad spectral feature between 2149 to 2153 eV in experiment during the second cycle of delithiation, especially the emerging higher energy peak at 2151 eV compared to the pristine Li$_3$PS$_4$. PDOS analysis shows that the low energy peak originates from the low-lying P-S antibonding π* ($p_z$) orbitals and the high energy doublet comes from the P-S antibonding σ* ($p_x$ and $p_y$) orbitals in the triangular local symmetry. Similar trends in spectral shape have been found in simulated organophosphorus spectra, where the 3-coordinated P sites show a larger split in the K-edge white line peak than 4-coordinated P sites[60].

Here we highlight the importance of the P 1*s* core level energy correction that includes the non-local correlation effect at the ground state. Without this correction, at the Kohn-Sham DFT level the spectrum of the PS$_3$ motif shifts to lower energy by 1.82 eV with respect to the PS$_4$ spectrum in pristine Li$_3$PS$_4$ as the dashed purple curve shown in Figure S8. As a result, this error in edge alignment deteriorates the comparison between simulation and experiment during the second cycle of delithiation.

Overall, the deformation of the PS$_4$ to PS$_3$ motif results in the π* and σ* splitting. The 1*s* → σ* transition corresponds to the emergence of the new peak at 2151 eV, while the 1s → π* transition slightly broadens the white line at lower energy.

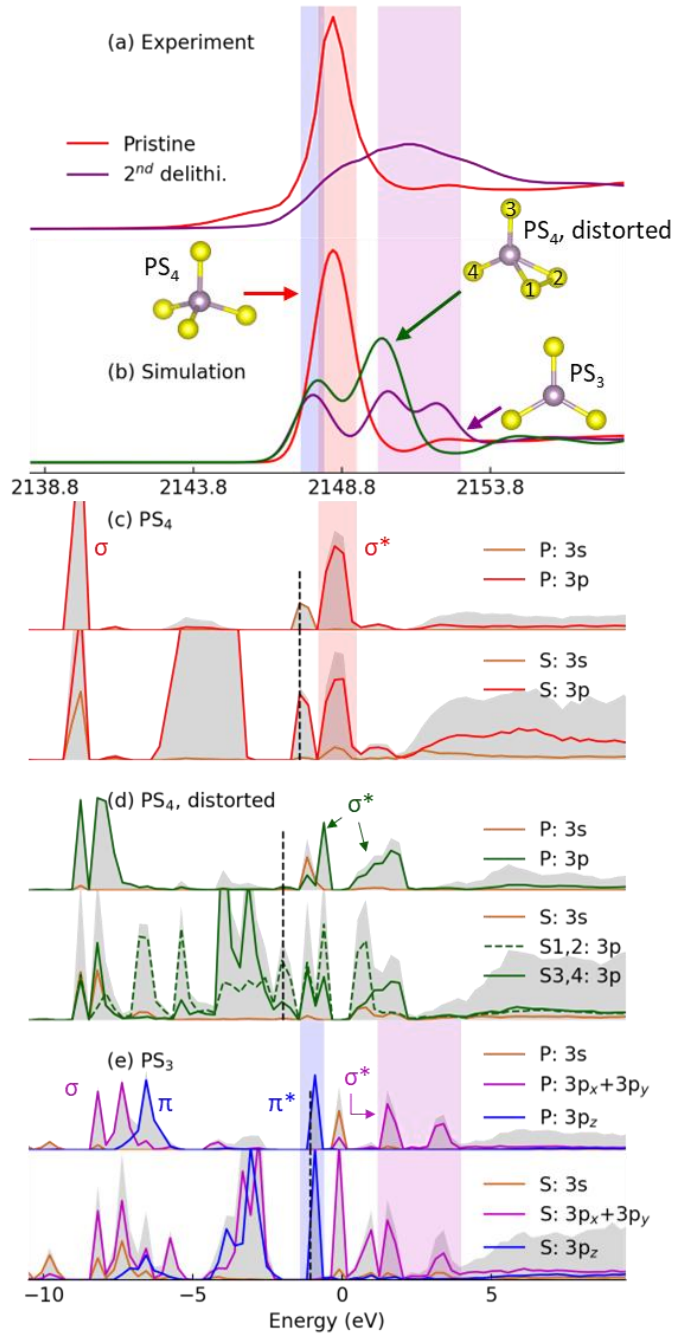

*Figure 5. (a) Experimental phosphorus XANES spectra. Error in the measured intensity is better than 0.1% and the photon energy better than 0.05 eV. (b) Simulated spectra of selected sites. The red, green, and purple curves in (b) are representative spectra of P sites with $PS_4$, distorted $PS_4$, and $PS_3$ motifs, respectively. (c – e) The corresponding projected density of states (PDOS) of the structures used in (b) calculated in the presence of a full core hole. (c) PDOS of 3s and 3p orbitals of phosphorus and sulfur in $PS_4$ motif. (d) PDOS of 3s and 3p orbitals of phosphorus and sulfur in distorted $PS_4$ motif. The green dashed and solid lines represent the PDOS of sulfur sites with and without S-S bond, respectively. (e) PDOS of 3s, $3p_x + 3p_y$, and $3p_z$ orbitals of phosphorus and sulfur in $PS_3$ motif. In the PDOS plots, the black dashed line indicates the position of the Fermi energy, and the grey shaded area represents the total DOS. The energy of the PDOS corresponds to the DFT energy levels with features aligned to the experimental spectra.*

*Clustering of S K-edge XANES*

Figure 6 shows the clustering results of simulated S K-edge XANES. Like the P XANES, the S XANES spectra were first aligned to the first major peak position and then clustered into six groups via the K-Means algorithm, as shown in Figure 6 (a-f). The six clusters are arranged in descending order by Li stoichiometry. Figure 6 (g-i) show the average values of structural descriptors within each cluster. Figure S5 shows the patterns from the structural PCA using the SOAP descriptor for all the S sites in the structural database.

In Cluster 1, the S sites have an average of three Li neighboring atoms and no S neighbors. These configurations also have relatively high Li stoichiometry, suggesting that these sulfur sites are from pristine and slightly delithiated β-LPS. It is important to note that the stoichiometry of a configuration is determined for the entire supercell as an average quantity and does not necessarily reflect the number of Li neighbors at a particular S site in that configuration, which is a local quantity. In a configuration with low Li stoichiometry, there could still be S sites with three or two Li neighbors.

Clusters 2 and 3 contain S sites with an average of two and one lithium neighbor(s), respectively. Like Cluster 1, these S sites have no S neighbors. The main difference in the spectra as compared to Cluster 1 is the intensity change at the valley between the first and the second peaks, at about 2473 eV. We have observed a similar feature in our experiments in the S0 and S1 spectra in Figure 2. The S0 and S1 spectra are re-plotted in Figure 6 (k) for easy comparison with simulated spectra. The slightly delithiated 3.4 V sample shows an increase in intensity at 2473 eV compared to the spectrum of pristine LPS, as indicated by the arrow.

The spectra of Cluster 4 and 5 exhibit different features from the first three clusters and show an increase in the intensity of the first major peak. This peak corresponds to the 2473 eV peak we observed in the experimental spectra after the first and the second delithiation. The energy location of this peak relative to the pristine LPS spectra is not reflected here because simulated spectra were aligned to the first major peak in Figure 6. The location of this peak can be better seen in Figure S7 with the energy-aligned spectra. The structural descriptors show that S-S bonds start to form (Figure 6 (h)). The distance between S atom and its neighboring P atom is larger in Cluster 5 than in Cluster 4, as shown in Figure 6 (i). This indicates that the formation of S-S bond weakens the S-P bond. Further analysis shows that some S atoms in Cluster 5 lose their P neighbor (Figure 6 (j)), which corresponds to the formation of $PS_3$ structure in the analysis of P spectra. We note that peak intensities in Cluster 4 and 5 are much higher than experimental spectra. This is because there are both S sites with and without S neighbor in one configuration. Cluster 4 and 5 only select S sites with S neighbors, while the experimental spectra are combinations of all S sites in the sample.

Cluster 6 seems to be an outlier at first glance, as it has the lowest Li stoichiometry but does not show any S-S bonds. This is also due to different S sites in a single configuration. S sites with a S-S bond are statistically more likely to occur in configurations with low Li stoichiometry, but this does not mean that every S site in these configurations has an S-S bond, and Cluster 6 includes S atoms without S neighbors but with Li neighbors.

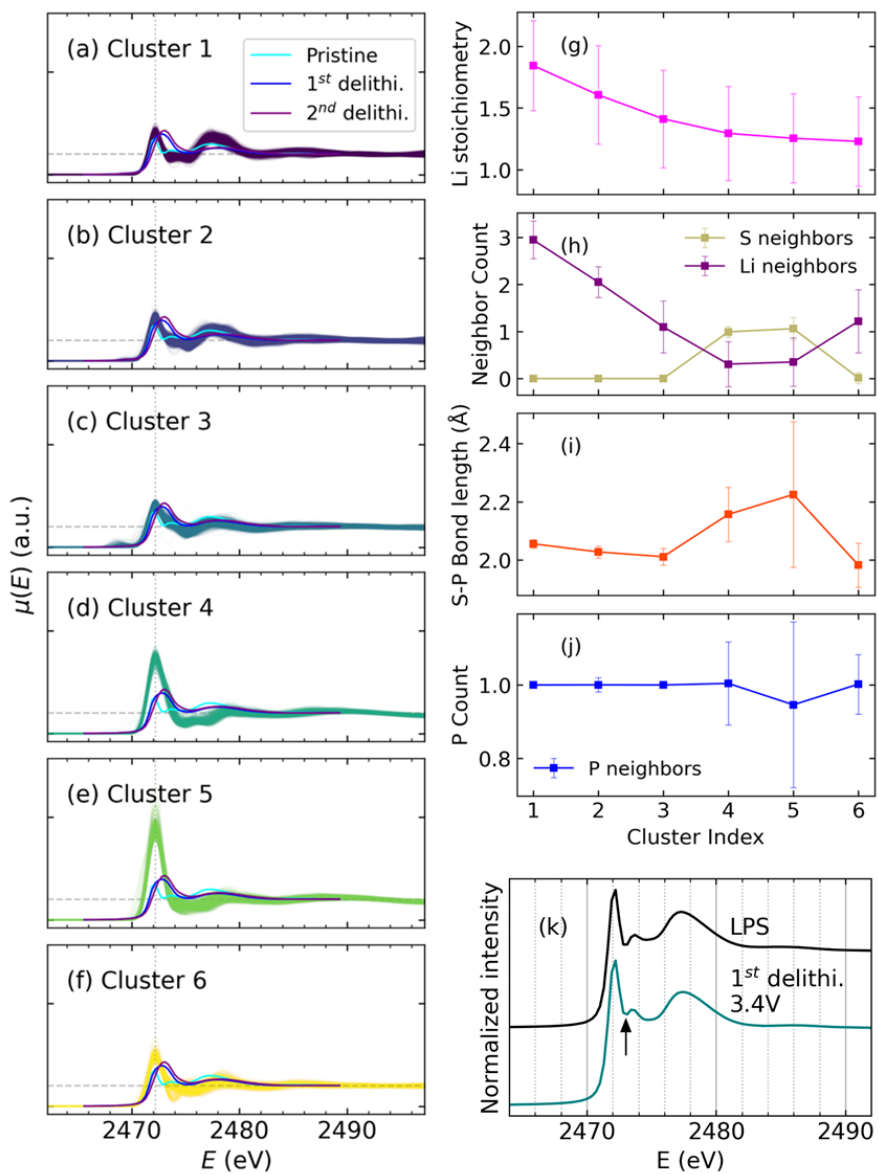

*Figure 6. Clustering of simulated S XANES spectra in the LPS dataset. The left column (a-f) plots the spectra in all six clusters, as well as experimental spectra measured on pristine LPS and LPS after the first and second delithiation. The right column (g-i) plots the averaged extracted structure descriptors in each of the 6 clusters with error bars representing the standard deviation. (g) Li stoichiometry, (h) the number of S and Li neighboring atoms, (i) S-P bond length, and (j) the number of neighboring P atoms. (k) Experimentally measured pristine and slightly delithiated LPS spectra. The arrow indicates the main difference between the two spectra at the valley between the first and the second peaks.*

### Atomic-scale insights into the delithiation of β-Li$_3$PS$_4$

To summarize our findings in data-driven P and S spectra analysis, we show that the delithiation of LPS starts with the decrease of Li neighbors around S atoms, which then leads to the formation of intra-tetrahedral S-S bonds accompanied by the large distortion of the PS$_4$ motif. In the second delithiation cycle, some P atoms lose their S neighbors and form PS$_3$ motifs.

By combining experimental XAS measurements, first-principles structure modeling and spectral simulations, and unsupervised ML, we can propose an atomic-scale model of the delithiation process of LPS. The onset of the delithiation is around 2.5 V (Figure 1, electrochemistry). The delithiation of LPS has three stages. In the first stage, the Li neighbors of S atoms decrease and the $PS_4$ structure deformation is minimal, as shown in Figure 7 (b). In the second stage, intra-tetrahedral S-S bonds start to form, and $PS_4$ structures are strongly distorted, as shown in Figure 7 (c). The intra-tetrahedral S-S bond gives rise to the observed new peak at 2473 eV in S K-edge XANES. In the third stage, $PS_4$ structures distort further, the minimum S-P-S bond angle is smaller than 70 degrees, and $PS_3$ motifs start to form, as shown in Figure 7 (d). The emerging broad peak at 2151 eV in P K-edge XANES can be attributed to the σ* band with large splitting in $PS_3$ and the highly distorted $PS_4$, as well as the π* band in $PS_3$. Structures such as the one in Figure 7 (c) can also exist at the third stage. The third stage only occurs in the second or following cycles, and further study is needed to reveal why the first cycle shows a different atomic-scale process. The CV curve and the spectral analysis suggest the following likely scenario. During the first lithiation process, the delithiated products are lithiated to a new phase at the LPS/C interface different from the original LPS. This new phase participates more actively than bulk LPS in the second and subsequent delithiation processes and forms the largely distorted $PS_4$ and $PS_3$ structures.

After the second cycle, the SSI layer continues to evolve. Figure S10 shows the P and S K-edge XANES data after eight cycles at 5 V. The P XANES data shows an increase in signal at 2152 eV, indicating an increase of distorted $PS_4$ structures compared to the second cycle. The peak energy is also about 1 eV higher than the peak position observed in Figure 2 (a), suggesting that the distortion of the $PS_4$ structure is more severe. The S XANES data at the eighth cycle is similar to that in the second cycle. The finding of $PS_4$ structure distortion is also supported by extended X-ray absorption fine structure (EXAFS) measurements. Figure S11 shows the EXAFS spectra of the P K-edge measured on LPS-C before and after cycling. In pristine LPS, the first shell of P atom corresponds to the P-S bonds with bond lengths of approximately 2.05 Å. After eight cycles, the intensity of the first shell decreases, indicating a disorder in the $PS_4$ tetrahedral structure.

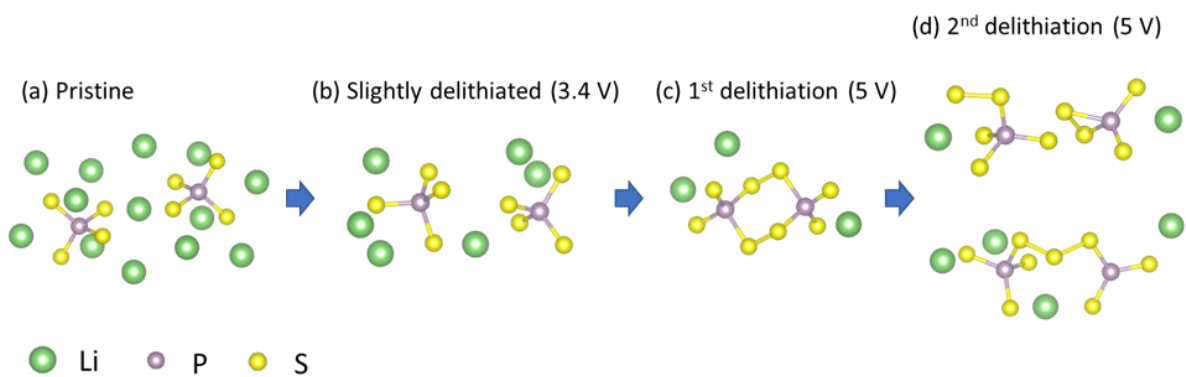

Figure 7. LPS configurations at different delithiation stages, selected from the structural database. (a) Pristine LPS. (b) S atoms lose their affinity Li neighbors, corresponding to the slightly delithiated LPS at 3.4 V during the first delithiation process. (c) S-S bond forms and $PS_4$ structure distorts, corresponding to the structure at the end of the first delithiation. (d) Structures after the second and following delithiation cycles. In the top structure, S atoms within the same $PS_4$ tetrahedron move closer to each other and form S-S bond, $PS_4$ largely distorts, forming a sharp S-P-S angle. In the bottom structure, $PS_3$ motif forms.

*Transport properties of SSI*

To elucidate how the changes in local structure affect the transport properties of the solid electrolyte, we measured the electrochemical impedance spectroscopy (EIS) at different charge stages and different cycle numbers. Figure S12 (a) shows that a highly resistive interphase layer forms after delithiation, and the change in interfacial resistance is reversible upon lithiation. This suggests that the Li-deficient interphase is a poor Li$^+$ conductor and adding Li can increase its ionic conductivity. The high resistance under the Li-deficient condition could be due to the formation of a large number of S-S bonds and the distortion of PS$_4$ tetrahedra, which increases the overall tortuosity of the Li$^+$ transport pathway[61]. The formation of this highly resistive interphase is detrimental to batteries' performance, leading to increased overpotential and limiting the charging capacity of the cell.

The increase of resistivity with cycle number (Figure S12 (b)) shows that this interphase layer does not passivate the LPS surface, and SSI grows with cycling, which is consistent with the XAS results in Figure S10. The poor passivation capability of the SSI suggests that this layer has finite electronic conductivity and still allows electron transport. A previous study[62] on sulfide electrolytes without electrochemical delithiation claimed that the formation of S-S bond decreases the band gap and increases the electronic conductivity of sulfide electrolytes. In our study, when delithiation is considered, S atoms are partially reduced to form a S$^{2-}$ and S$^-$ mixture as supported by the Bader charge analysis. The DOS plots of randomly selected structures with different lithium compositions are shown in Figure S13, where the DOS has a finite value at the Fermi level, confirming that the delithiated compounds become metallic. Our findings suggest that the delithiation process converts the electrolyte into a mixed electronic and ionic conductor[63] and facilitates uncontrollable SSI growth.

*Generalization of the spectral fingerprints of LPS to Li$_6$PS$_5$Cl*

Our study of LPS electrolytes established the spectral fingerprint of distorted PS$_4$ motifs. In particular, the main peak of P K-edge XANES is broadened and shifted to higher energy by about 2.5 eV in the delithiated LPS as compared to the pristine LPS. This spectral fingerprint is characteristic of the intra-PS$_4$ tetrahedral S-S bond and the formation of the PS$_3$ motif. This spectral fingerprint is a proxy of the chemical stability of phosphorus sulfides under the kinetic-driven delithiation process. We now extend the scope to an argyrodite-structured electrolyte, Li$_6$PS$_5$Cl (LPSCl). Argyrodite electrolytes have been found to have higher ionic conductivity than the basic LPS electrolyte[64], and are believed to form more stable interphases[44]. We note that although some of the S sites are substituted by halide ions, the tetrahedra around P atoms, i.e., 16e sites in the argyrodite structure, remain fully occupied by S$^{2-}$ anions[65]. Thus, we correlate this spectral fingerprint to the evolution of the PS$_4$ motif in LPSCl, which underscores the wide applicability of our results to different types of sulfide electrolytes with similar microstructure motifs.

We measured P and S K-edge spectra of the LPSCl-carbon composite after five cycles, as shown in Figure S14. The S K-edge spectra of LPSCl are similar to those of LPS in both pristine and cycled samples. A peak at 2473 eV appears after delithiation, indicating the formation of S-S bond. The P K-edge spectrum of the pristine LPSCl sample is similar to that of LPS due to the similar local structure. However, the P spectrum of cycled LPSCl exhibits a significant difference from cycled LPS. In LPSCl, the intensity of the white line peak at 2148.5 eV decreases and the peak broadens, without forming new peaks. Unlike LPS, the distortion of the PS$_4$ tetrahedra in LPSCl is greatly alleviated, which implies the Cl substitution can stabilize the local PS$_4$ structure. The results of LPSCl suggest that the origin of better stability of argyrodite electrolytes may come from the more stable $PS_4$ structure, while the impact of halide substitution is subject to further investigation.

**Conclusion**

In summary, we have unraveled the atomic-scale delithiation process of LPS electrolyte using XAS and data-driven spectra interpretation. The delithiation starts with a reduced number of Li neighbors that are bonded to S atoms, and then intra-tetrahedral S-S bond forms along with the distortion of the $PS_4$ tetrahedra. In the second and following cycles, the $PS_4$ structures are strongly distorted with sharp S-P-S angles and $PS_3$ structures start to emerge.

To our knowledge, this study is the first to explain the changes and the origin of new features in the XANES spectra of LPS material during delithiation with atomic-scale insights. Our results have several implications which can largely benefit the design and optimization of sulfide electrolytes. First, the distortion of the $PS_4$ tetrahedra and the formation of S-S bonds could be the origin of the large interfacial resistance of LPS. Second, the structural fingerprints obtained in this study can be used in the interpretation of other sulfide electrolytes with similar structures, e.g., the argyrodite electrolytes $Li_6PS_5X$ (X = Cl, Br, and I). The better stability of LPSCl over LPS could be due to a more stable $PS_4$ structure. Third, the analysis of XANES data is always intriguing, and the developed methodology for XAS data interpretation is applicable to the analysis of XANES data in other areas.

**Methods:**

*Cell assembly and electrochemistry*

LPS-C composite was made by mixing $Li_3PS_4$ power (NEI Corporation) with carbon black at a ratio of 7:3 by mass. The LPS material was characterized using scanning electron microscopy (SEM), XPS, and XRD, shown in Figure S1 – S3. The mixed powder was ball milled at 400 rpm at room temperature under an argon atmosphere. To avoid overheating, the ball milling was conducted with 10 min of milling and 15 min of rest. This process was repeated 40 times, so the total time of milling was 400 min. The purpose of using LPS-C composite is to increase the surface-to-volume ratio and to promote the surface redox reactions of LPS.

The electrolyte pellet was fabricated by pressing 60 mg of LPS in a 1 cm diameter die at 280 MPa. The thickness of the pellet was about 400 µm. 10 mg LPS-C composite powder was then spread over the top of the LPS pellet and pressed at 380 MPa together with the LPS pellet so that the working electrode (LPS-C composite) and the electrolyte (LPS) were in close contact. The counter/reference electrode was Li-In alloy. The LPS-C | LPS pellet and Li-In foil were transferred into a pressure-controlled split cell (MTI Corporation) and operated at a stack pressure of approximately 4 MPa. A schematic of the cell configuration is shown in Figure 1 (a).

LPSCl-C composites were fabricated and cycled in the same way as LPS using in-house synthesized LPSCl powder. The synthesis of LPSCl was performed in an argon filled glovebox. Stoichiometric amounts of LiCl (Lithium Chloride 99%, Strem Chemicals), $Li_2S$ (Lithium Sulfide 98%, Strem Chemicals), and $P_2S_5$ (Phosphorus Pentasulfide 99%, Sigma Aldrich) were ball milled at 400 rpm for 18 hours, then loaded into a capped quartz crucible and heated to 550°C at 90℃/hour, and held at 550°C for 12 hours. The product was then cooled for 6 hours to room temperature and ground with a mortar and pestle.

The electrochemistry tests were conducted with a BioLogic potentiostat VSP-300. All tests were done in an argon-filled glovebox. Cyclic voltammetry (CV) experiments were performed at 0.1 mV/s with 0 V to 5 V voltage range. The CV measurements started with anodic sweep from open-circuit voltage (OCV) to 5 V, followed by cathodic sweep from 5 V to 0 V. The cells were disassembled at different stages in the CV measurements for characterization. All voltages in this manuscript are reported with respect to the $Li^+$/Li redox potential. Error in voltage and current measurements are better than 1 mV and 1 μA respectively.

*XAS Characterization*

XAS experiments of P and S K-edges were performed in fluorescence yield (FY) mode at the 8-BM (TES) beamline, and in electron yield (EY) mode at the 7-ID-2 (SST-2, operated by the National Institute of Standards and Technology) beamline at the National Synchrotron Light Source II (NSLS-II). The beam spot sizes are 2.5 mm × 5 μm at TES, and 1 mm × 1 mm at SST-2, respectively. For FY experiments at the TES beamline, samples were sealed between Kapton tape and a 5 μm thick polypropylene (PP) film in an argon-filled glovebox with the LPS-C side facing the PP film. The samples were then sealed in aluminized polymer pouch and transferred to the helium chamber at the TES beamline for XAS measurements. For EY experiments at the SST-2 beamline, samples were mounted on a sample bar and sealed in aluminized polymer pouch in the glovebox, and then transferred to the measurement chamber through an argon-filled glovebag. XAS measurements were performed in the ultra-high vacuum (UHV) chamber at SST-2 beamline. The LPS-C side of the sample was electrically connected to the sample bar with copper tape, and the Li-In side was covered with Kapton tape for insulation; the sample bar was grounded. XAS spectra were processed using the Athena software package[66]. From the signal-to-noise ratio of the measured data, error in the XAS intensity is expected to be better than 0.1%. Error in photon energy calibration is better than 0.05 eV.

*Computational delithiation*

All DFT[67, 68] calculations were performed within the projector-augmented wave (PAW) method[69, 70] as implemented in the Vienna ab initio Simulation Package (VASP)[71-73]. To generate atomic models of delithiated LPS, a computational delithiation of a LPS supercell with composition $Li_{12}P_4S_{16}$ was carried out by enumerating possible lithium/vacancy orderings with the *Enumlib* software[74-76] and Pymatgen[77]. *B*-LPS structure is orthorhombic with the space group Pnma (62), and lithium atoms occupy sites 8d, 4b and 4c with occupation probabilities of 1.0, 0.70, and 0.30[78]. The computational delithiation was performed with two approaches: First, we generated all Li/vacancy orderings independently based on the initial fully lithiated $Li_{12}P_4S_{16}$ structure, which was followed by an optimization of the atomic positions and lattice parameters. This approach mimics the situation where the material at each lithium content is in thermodynamic equilibrium. We then repeated the delithiation a second time sequentially, such that in each step one more Li atom was removed from the relaxed structure, approximating a kinetically controlled lithium extraction. The sequential delithiation visits a redundant subset of the structures considered in the independent delithiation, except for low lithium contents where strong structural rearrangements were observed in the sequential delithiation that were not captured by the independent enumeration. A total of 660 atomic configurations were obtained, 591 from independent and 69 from sequential enumeration.

The spin-polarized Perdew, Burke and Ernzerhof (PBE) exchange-correlation functional under the generalized gradient approximation[79] was used for structure optimization. The plane-wave cut-off energy of 520 eV and a Gamma-centered reciprocal space discretization of 25 $k$-points per Å$^{-1}$ were chosen. The electronic convergence criterion was set to 10$^{-5}$ eV. The β-Li$_3$PS$_4$ crystal structure was obtained from the Materials Project database[80] (materials id: mp-985583). DFT calculations for the 660 enumerated structures were performed, and the formation energies were calculated with respect to the reference phases Li$_3$PS$_4$ and PS$_4$ according to the equation:

$$E_f(\text{Li}_{3-3x}\text{PS}_4) = E(\text{Li}_{3-3x}\text{PS}_4) - (1-x)\,E(\text{Li}_3\text{PS}_4) - x\,E(\text{PS}_4),$$

where $E$ is the DFT energy. The thermodynamically stable delithiated phases were determined via a convex-hull construction[81], and the equilibrium voltage between each two stable phases was calculated as

$$V(x_1, x_2) = -\frac{E(\text{Li}_{x_1}\text{PS}_4) - E(\text{Li}_{x_2}\text{PS}_4) - (x_1 - x_2)E(\text{Li})}{(x_1 - x_2)\,F},\, x_1 > x_2,$$

where F is the Faraday constant. The oxidation states of Li, S, and P atoms were assigned based on a Bader charge analysis[82].

### S and P K-edge XANES Spectral simulation

The S and P K-edge XANES simulations were performed using the core-hole potential method[83] implemented in VASP 6.2.1[84]. The many-body final state effects are treated by the Kohn-Sham (KS) self-consistent solution with the PBE functional at the presence of a full core hole. Supercells of size no smaller than 9 Å along each principal axis are constructed to minimize the spurious interactions from image cells. The GW type of the projector augmented wave (PAW) pseudopotentials were used to obtain more accurate description of the post-edge region[55]. To yield a better comparison with experimental spectra, the simulated spectra were broadened using Gaussian function with a standard deviation of σ = 0.6 eV.

Edge alignment is important for comparing XANES spectra calculated on symmetrically inequivalent sites of materials in the structure database. Different edge alignment algorithms are implemented in different codes, and each has its limitations. A systematic benchmark of the edge alignment procedure against experimental standards is lacking[85]. We first compute the excitation energy under the single particle picture using KS DFT energy levels according to $\Delta\varepsilon_{ck} = \varepsilon_{ck} - \varepsilon_{1s}$, where $\varepsilon_{ck}$ is the energy level of a conduction band ($c$) at wavevector $\boldsymbol{k}$ in the Brillouin zone at the presence of a full core hole (i.e., the final state) and $\varepsilon_{1s}$ is the 1s core level energy of the neutral absorber atom of the ground state (i.e., the initial state). From the relative edge alignment, the absolute edge alignment can be obtained by aligning a reference spectrum to the corresponding experimental standard with a constant energy shift, and the constant is applied to all the simulated spectra in the database. In this study, we chose β-LPS as the reference system.

The edge alignment procedure described above is subject to the error in KS orbital energy levels due to the approximation in the local or semi-local exchange-correlation potential. At the ground state the top of the valence band of LPS and delithiated LPS models is dominated by S *3p* states, and the bottom of the conduction bands is derived from the P-S $\pi^*$ or $\sigma^*$ antibonding orbitals of the P polyhedra (see more detailed discussion in the Results section under "Clustering of P K-edge XANES"). As a result, while the correlation effects of the S 1s orbital is mostly local and can be captured by semi-local DFT, the correlation effects of the P 1s orbital is non-local with significant contributions from the dielectric screening of neighboring S atoms through the S *3p* to P-S $\pi^*$ and $\sigma^*$ transitions. Therefore, the edge alignment of the P K-edge XANES based on KS orbital energy levels requires further correction.

A widely used edge alignment correction procedure in the core-hole pseudopotential calculations replaces the energy difference between KS energy levels by the total energy difference between the final state and initial state[86, 87]. In particular, the onset of the X-ray absorption is approximated by placing a 1s electron of the absorber atom at the bottom of the conduction band and leaving a full 1s core hole behind. However, this method cannot be applied to the delithiated systems, where delithiation creates holes at the valence band top of β-LPS. Then in the total energy calculation of the excited system, the core electron will be placed at the top of the valence band to fill the holes and cause a misalignment. An alternative is to perform the self-energy correction for the 1s core level[88] and the conduction band orbitals under the GW approximation[89]. However, GW calculations are very expensive and not feasible for our system. In this study, we consider that the dominating edge alignment correction in P K-edge comes from the static non-local correlation effects of the P 1s orbital, which can be calculated with the OCEAN code [86, 87], accounting for the adiabatic relaxation of the system in response to the removal of a core electron[90]. In OCEAN calculations, we set the "screening radius" keyword to 6 Bohr and dielectric macroscopic constant to the value of β-LPS at 3.873 calculated using the density functional perturbation theory[91] implemented in the Quantum ESPRESSO code[92]. Finally simulated P K-edge XANES site spectra were aligned relatively to each other according to $\Delta \varepsilon_{c\boldsymbol{k}}^P = \varepsilon_{c\boldsymbol{k}}^P - \varepsilon_{1s}^P + \varepsilon_{screen}^P$, where $\varepsilon_{screen}^P$ is the adiabatic static screening term calculated from OCEAN. This current core-level correction scheme does not consider the intra-band screening effects, which is a subject for future investigation.

*Machine learning*

To assist in our analysis, K-means clustering was used to classify the simulated spectra into different clusters. The algorithm was implemented using the scikit-learn package[93]. We used cluster number *k* = 5 for P K-edge XANES and *k* = 6 for S K-edge XANES spectra. The number of clusters, *k*, is an input to the K-means clustering algorithm. In practice, an optimal choice of the number of clusters depends on the problem to address. In this study, because our motivation is to use the clustering algorithm to identify structural descriptors to interpret the spectra, the choice was made to establish a consistent correlation between local chemical environment (e.g., neighboring atoms and chemical bonding) and distinct spectral clusters. In addition, the numbers of clusters were corroborated using Elbow criterion[94] as explained in "K-Means parameters and determining the number of clusters" section in the SI by inspecting the inter-cluster distances against the number of clusters (see Figure S15). More details about the parameters used in K-Means clusters can also be found in the SI.

**Data availability**

The simulated phosphorus and sulfur K-edge XAS data, as well as the input files for spectra calculation using VASP implementation, have been deposited in the Materials Cloud database[95].

## Acknowledgments

This work was funded in part by the U.S. Department of Energy (DOE) Office of Energy Efficiency and Renewable Energy, Vehicle Technologies Office, Contract No. DE-SC0012704. This research used 8-BM, 7-ID-2, and 28-ID-2 of the National Synchrotron Light Source II, and the theory and computational facility and Proximal Probes Facility of the Center for Functional Nanomaterials (CFN), which are U.S. Department of Energy (DOE) Office of Science User Facilities operated for the DOE Office of Science by Brookhaven National Laboratory under Contract No. DE-SC0012704. Beamline 7-ID-2 is funded and operated by the National Institute of Standards and Technology. We also acknowledge computing resources from Columbia University's Shared Research Computing Facility project, which is supported by NIH Research Facility Improvement Grant 1G20RR030893-01, and associated funds from the New York State Empire State Development, Division of Science Technology and Innovation (NYSTAR) Contract C090171, both awarded April 15, 2010. The research used the Scientific Data and Computing Center at Brookhaven National Laboratory under Contract No. DE-SC0012704. We thank Yusuf Celebi for his technical support. D.L. thanks Dr. John Vinson for helpful discussions.

## Author contributions

C.C., S.Y., N.A., A.U., D.L., and F.W. conceived the idea and designed the project. C.K., J.S.S., H.G., and A.U. performed structural calculations and analysis. M.R.C. performed the spectra calculations. M.R.C. and C.C. performed machine learning analysis. K.S. and D.S. performed the synthesis of argyrodite electrolytes. C.C. performed the electrochemistry measurements. S.L. and K.C. performed the XRD and SEM measurements. C.C., S.-M.B., Y.D., and C.W. performed XAS measurements. X.T. performed XPS measurements. C.C., M.R.C., A.U., and D.L. co-wrote the manuscript. All authors discussed the results and contributed to the manuscript.

## Declaration of interests

The authors declare no competing interests.

**Disclaimer:** Commercial equipment, instruments, or materials are identified in this paper to specify the experimental procedure adequately. Such identification is not intended to imply recommendation or endorsement by the National Institute of Standards and Technology, nor is it intended to imply that the materials or equipment identified are necessarily the best available for the purpose.

# Supporting Information

## Atomic Insights into the Oxidative Degradation Mechanisms of Sulfide Solid Electrolytes

**Authors:**

Chuntian Cao[1,†], Matthew R. Carbone[1,†], Cem Komurcuoglu[2], Jagriti S. Shekhawat[2], Kerry Sun[2], Haoyue Guo[2], Sizhan Liu[3], Ke Chen[3], Seong-Min Bak[4,5], Yonghua Du[4], Conan Weiland[6], Xiao Tong[7], Dan Steingart[2,8,9], Shinjae Yoo[1], Nongnuch Artrith[10], Alexander Urban[2,8,*], Deyu Lu[7,*], Feng Wang[3,*,a]

[1] Computational Science Initiative, Brookhaven National Laboratory, Upton, NY 11973, USA

[2] Department of Chemical Engineering, Columbia University, New York, NY 10027, USA

[3] Interdisciplinary Science Department, Brookhaven National Laboratory, Upton, NY 11973, USA

[4] National Synchrotron Light Source II, Brookhaven National Laboratory, Upton, NY 11973, USA

[5] Department of Materials Science and Engineering, Yonsei University, Seoul 03722, Republic of Korea

[6] Material Measurement Laboratory, National Institute of Standards and Technology, Gaithersburg, MD 20899, USA

[7] Center for Functional Nanomaterials, Brookhaven National Laboratory, Upton, NY 11973, USA

[8] Columbia Electrochemical Energy Center (CEEC), Columbia University, New York, NY 10027, USA

[9] Department of Earth and Environmental Engineering, Columbia University, New York, NY 10027, USA

[10] Materials Chemistry and Catalysis, Debye Institute for Nanomaterials Science, Utrecht University, 3584 CG Utrecht, The Netherlands

Equal contribution: †

[a] Present address: Applied Materials Division, Argonne National Laboratory, 9700 S. Cass Avenue, Lemont, IL 60439

1. **Characterizations of the Li₃PS₄ material**

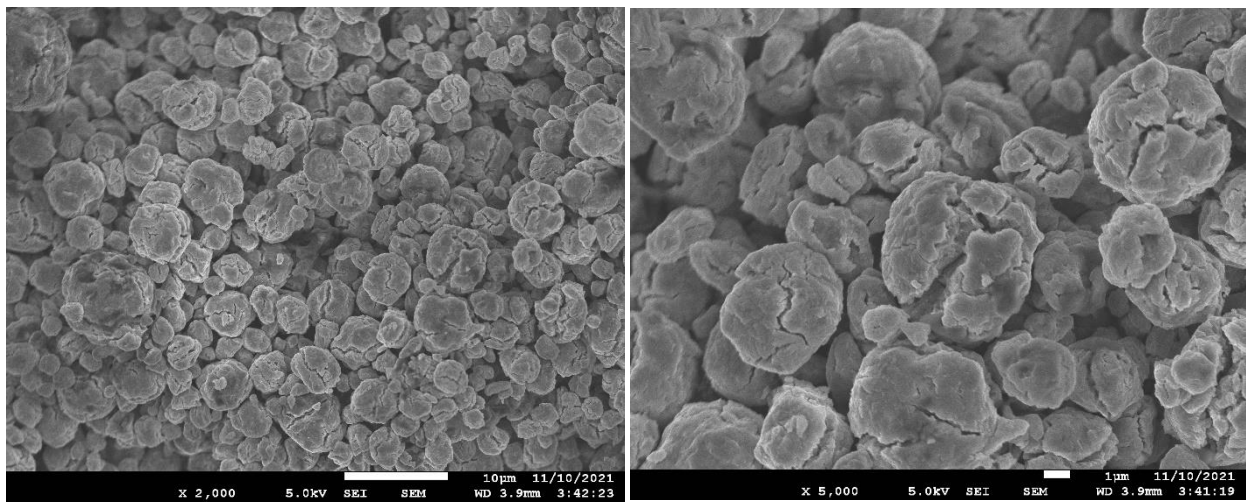

*Figure S1. Scanning electron microscopy (SEM) images of Li$_3$PS$_4$ particles from NEI corporation. The measurement was conducted at the Center for Functional Nanomaterials (CFN) at Brookhaven National Laboratory.*

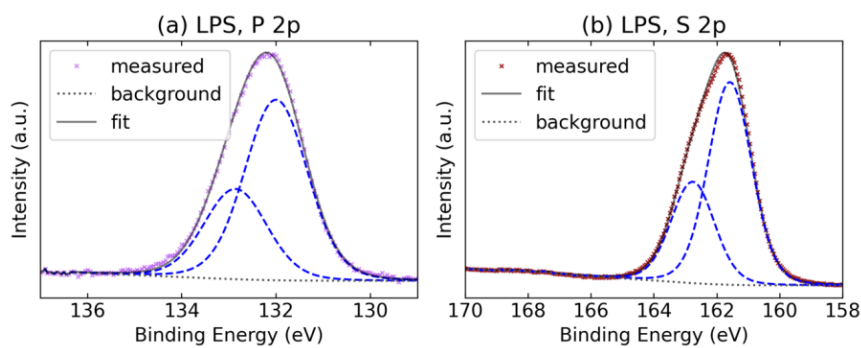

*Figure S2. X-ray photoelectron spectroscopy (XPS) data and fitting of pristine Li$_3$PS$_4$. (a) P 2p, (b) S 2p. The result is consistent with that previously reported in literature[1, 2]. The blue dashed lines in each figure indicate the fitting components, 2p$_{1/2}$ and 2p$_{3/2}$. XPS measurement was conducted at the Center for Functional Nanomaterials (CFN) at Brookhaven National Laboratory. The X-ray source is Al K-alpha at 1486.6 eV, and the pass energy is 25 eV. XPS fitting was performed using CasaXPS[3]. From the signal to noise, error in the measured intensity is expected to be better than 0.1% and in the binding energy better than 50 meV.*

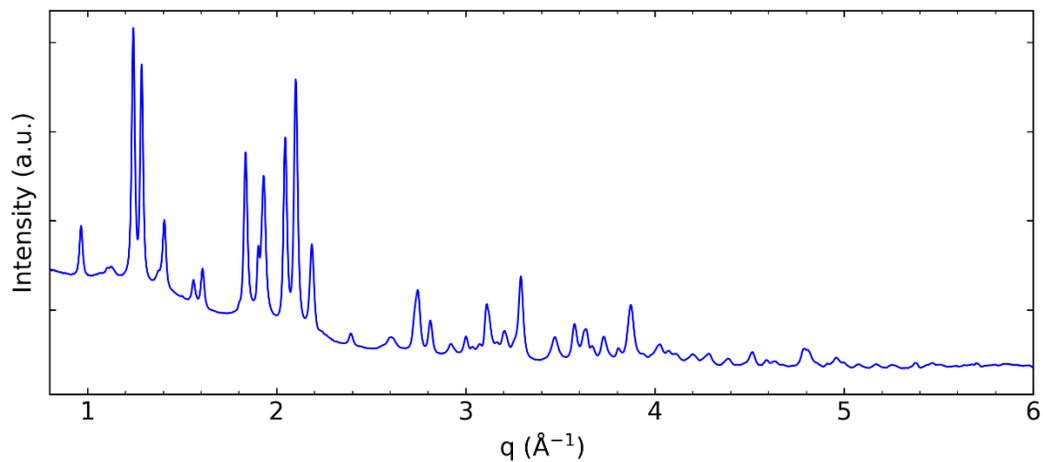

*Figure S3. X-ray diffraction (XRD) characterization of the $Li_3PS_4$ material from NEI corporation. The XRD pattern is consistent with previously reported in literature[4]. The experiment was conducted at beamline 28-ID-2 at NSLS-II.*

## 2. Smooth Overlap of Atomic Positions (SOAP) analysis in each cluster

We perform complementary analyses of the structures in each cluster using Smooth Overlap of Atomic Positions (SOAP) descriptors[5]. In Figure S4, we compute the SOAP descriptors for each absorbing P site, and then use Principal Component Analysis (PCA) to decompose those descriptors into two dimensions. Figure S5 shows the patterns from the structural PCA using the SOAP descriptor for all the S sites in the structural database. The clustering patterns in the "structural space" clearly mirror the changes in the spectral space, confirming our hypothesis that indeed it is the local structural changes that drive the changes in the XANES spectra.

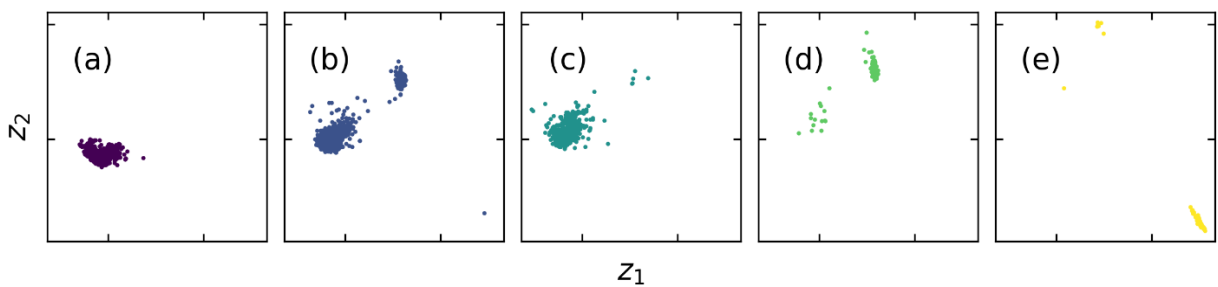

Figure S4. The corresponding patterns from the structural PCA using the SOAP descriptor for all the P sites in the structural database. (a – e) correspond to Cluster 1 – 5 in Figure 4 in the main text.

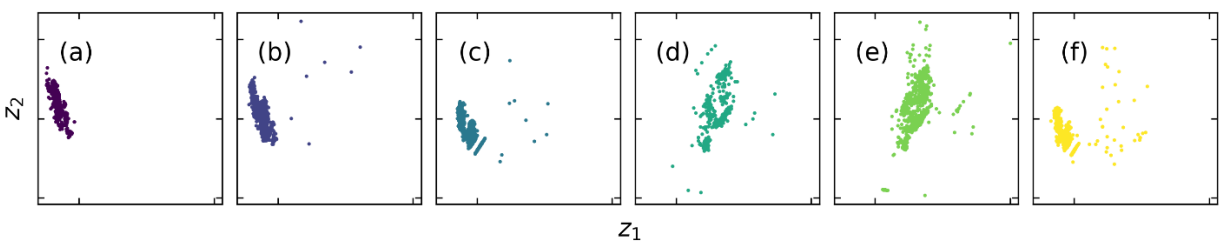

Figure S5. The corresponding patterns from the structural PCA using the SOAP descriptor for all the S sites in the structural database. (a – f) correspond to Cluster 1 – 6 in Figure 6 in the main text.

## 3. Plots of energy-aligned simulated P and S spectra in each cluster

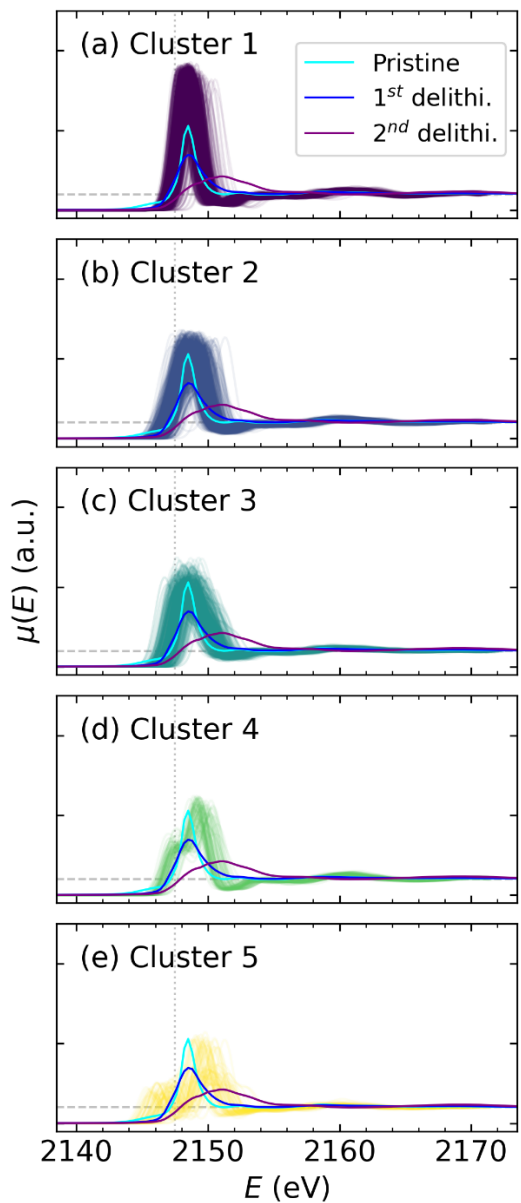

*Figure S6. Energy aligned simulated phosphorus spectra. The data is the same as Figure 4 in the main text, except that the energy is aligned using the method stated in the Methods section in the main text.*

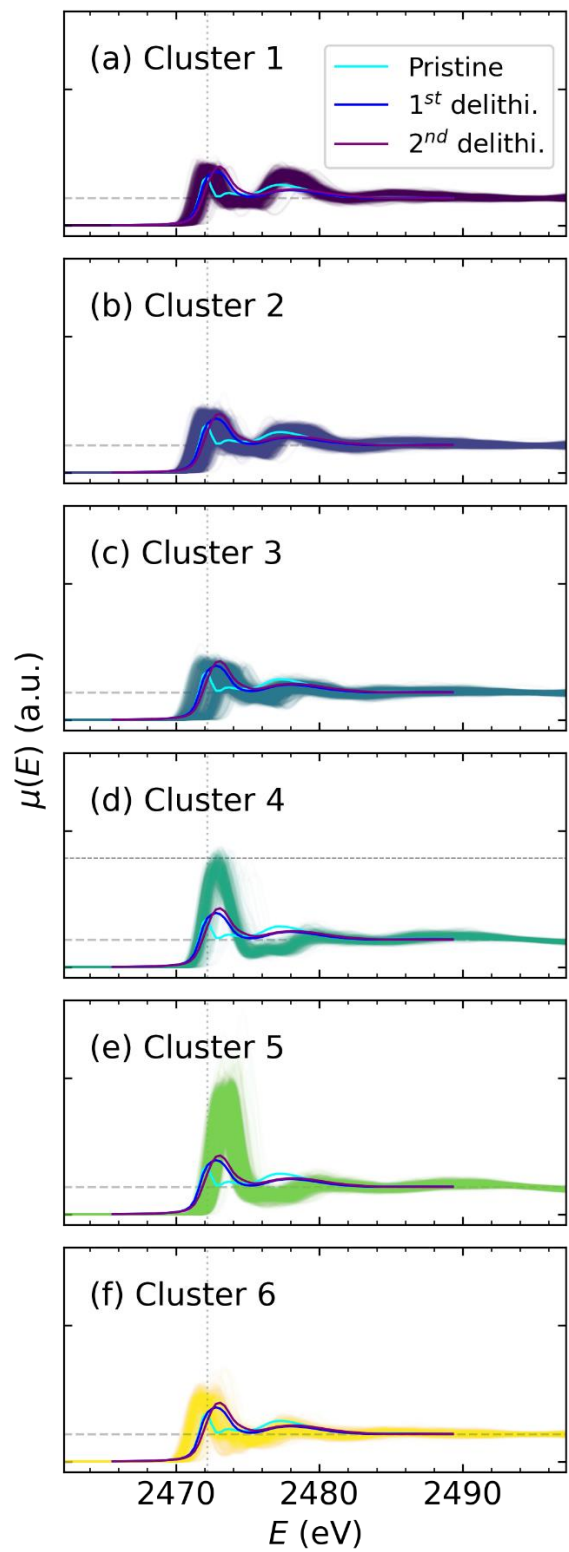

*Figure S7. Energy aligned simulated phosphorus spectra. The data is the same as Figure 6 in the main text, except that the energy is aligned using the method stated in the Methods section in the main text.*

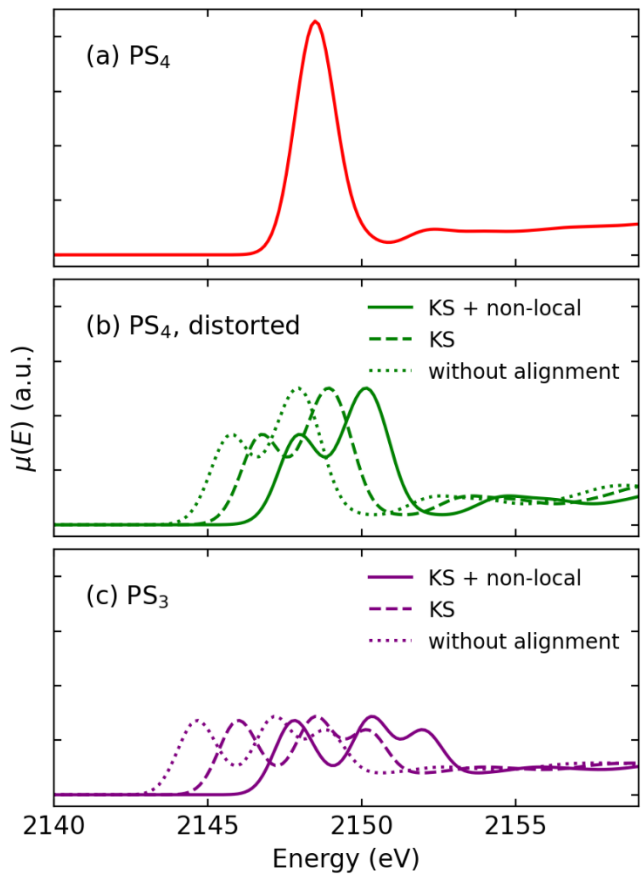

*Figure S8. Simulated P K-edge spectra of selected sites with representative structural motifs with and without edge alignment. (a) Ideal $PS_4$ motif. (b) distorted $PS_4$ motif. (c) $PS_3$ motif. In (b) and (c), the dotted lines show the spectra without alignment, the dashed lines show the spectra with only KS correction, and the solid lines show the spectra with both KS correction and non-local correction effect.*

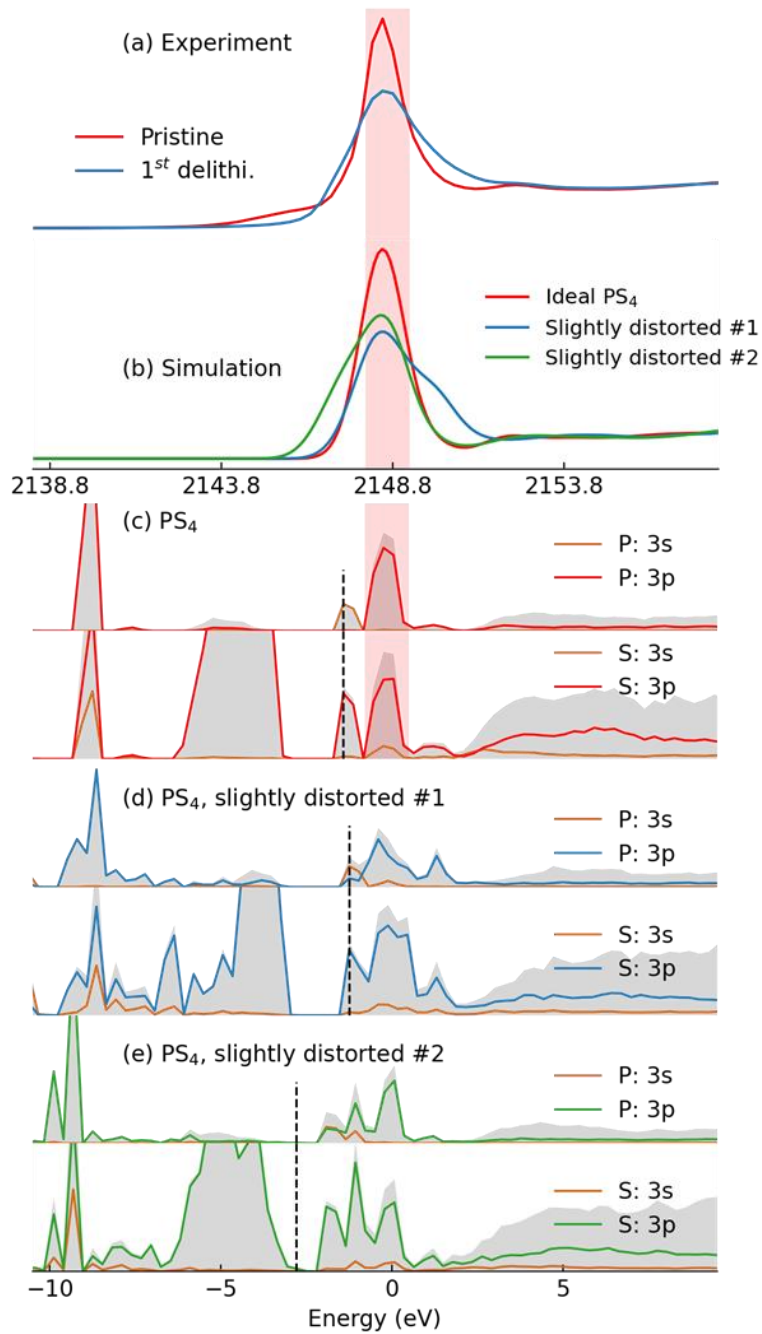

*Figure S9. (a) Experiment phosphorus XANES spectra. (b) Simulated spectra of selected sites, with red curve from ideal $PS_4$ tetrahedron, and the blue and green curves from slightly distorted $PS_4$. (c – e) The corresponding projected density of states (PDOS) plots calculated from the same structures of the simulated spectra in (b). In the PDOS plots, the black dashed line indicates the position of the Fermi energy, and the grey shaded area represents the total DOS. Error in the measured XANES spectra is expected to be better than 0.1% and in the photon energy better than 0.05 eV.*

## 4. XANES and EXAFS data after multiple delithiation cycles

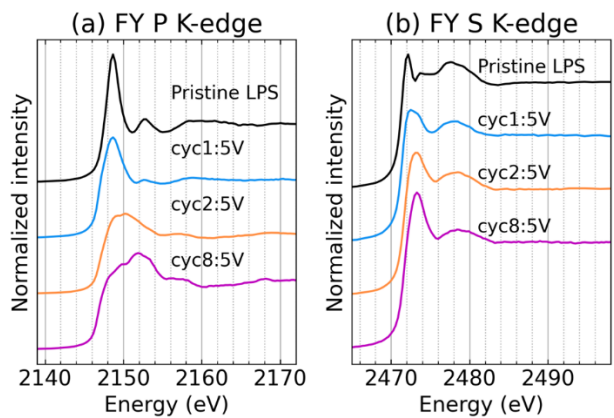

*Figure S10. (a) P K-edge and (b) S K-edge XANES data on delithiated LPS-C composite after different cycles. The data were collected in fluorescence yield mode at TES beamline at NSLS-II. Pristine LPS, cyc1-5V, and cyc2-5V data are also shown in Figure 2 (c) and (d) in the main text. Error in the measured XANES spectra is expected to be better than 0.1% and in the photon energy better than 0.05 eV.*

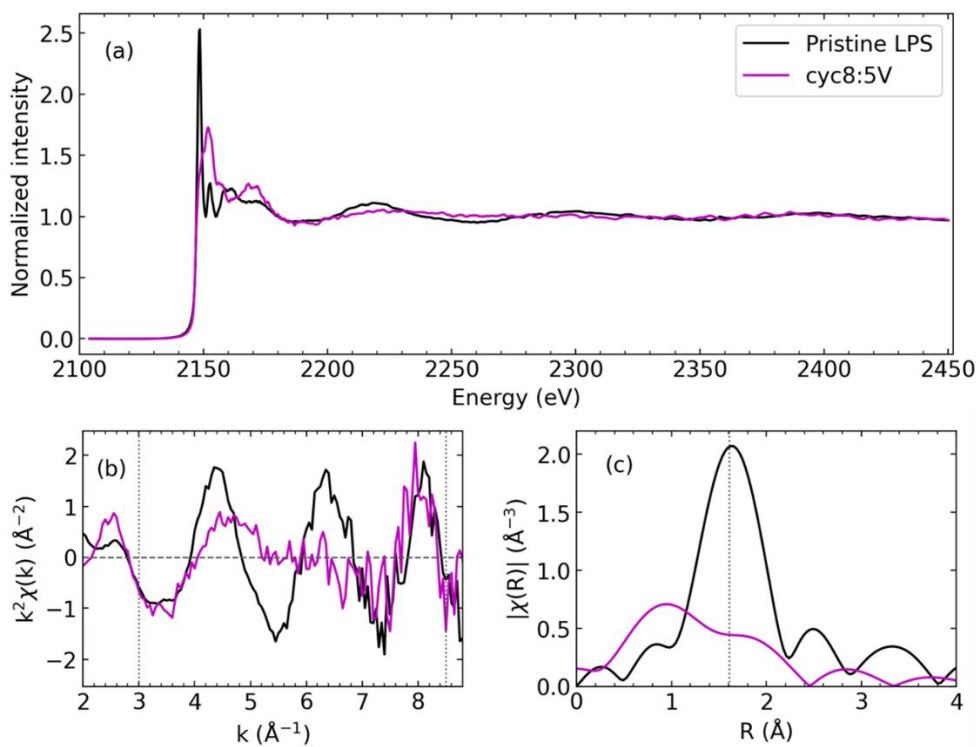

*Figure S11. (a) P K-edge XAS data on LPS-C composite before and after cycling. (b) k-space and (c) R-space extended X-ray absorption fine structure (EXAFS) of P K-edge. Black curve: pristine LPS; magenta curve: after eight cycles at the delithiation state. The Fourier transform range of k is 3 Å$^{-1}$ to 8.5 Å$^{-1}$. Data were measured at TES beamline at NSLS-II in fluorescence mode. Error in the measured XANES spectra is expected to be better than 0.1% and in the photon energy better than 0.05 eV.*

## 5. Characterize interface resistance with impedance spectroscopy

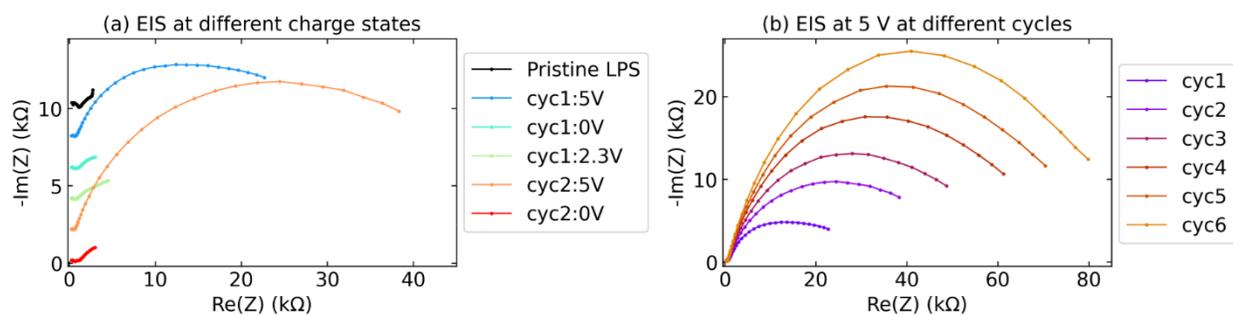

*Figure S12. Electrochemical impedance spectroscopy (EIS) of LPS-C | LPS |Li-In cell. (a) EIS measured at different charge states during the first two cycles. (b) EIS measured at delithiation state (5 V vs. Li$^+$/Li) at different cycles.*

## 6. Density of states plots of delithiated compounds

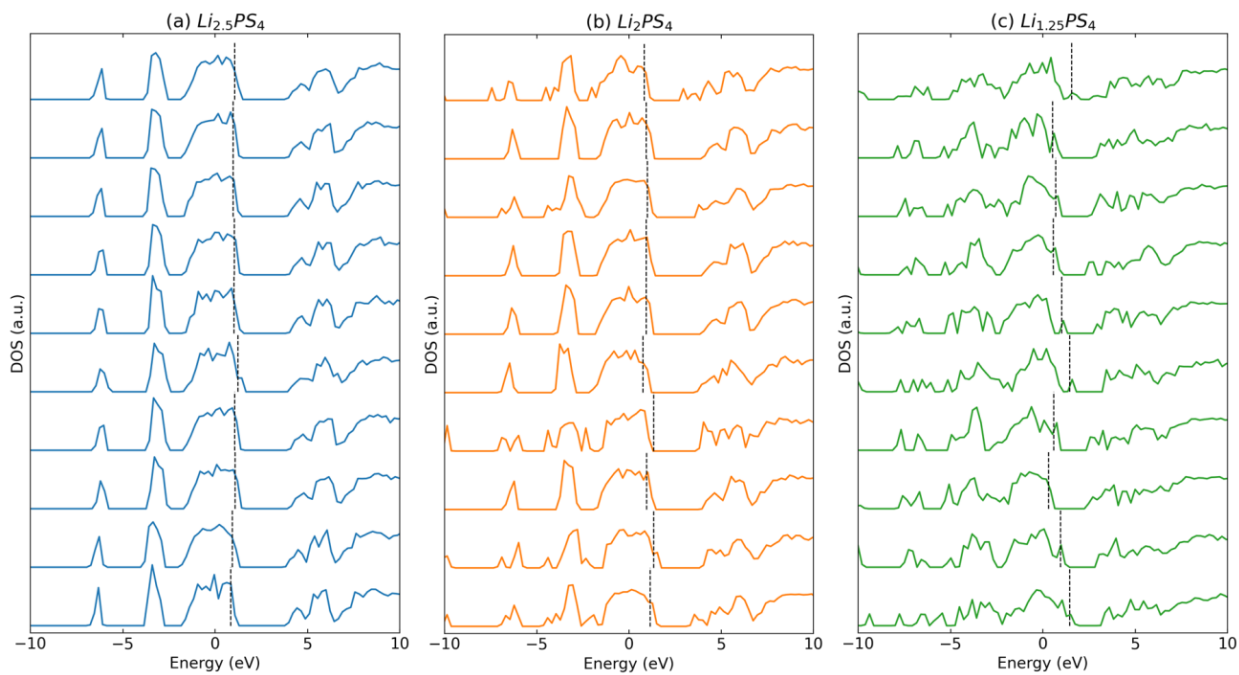

*Figure S13. The total density of states (DOS) plots of delithiated Li$_x$PS$_4$ compounds. Configurations from three delithiation compositions were selected, (a) Li$_{2.5}$PS$_4$, (b) Li$_2$PS$_4$, and (c) Li$_{1.25}$PS$_4$. In each subplot, ten compounds were randomly selected and plotted. The black dashed vertical line indicates Fermi energy level.*

## 7. Delithiation of argyrodite structured Li$_6$PS$_5$Cl delithiation

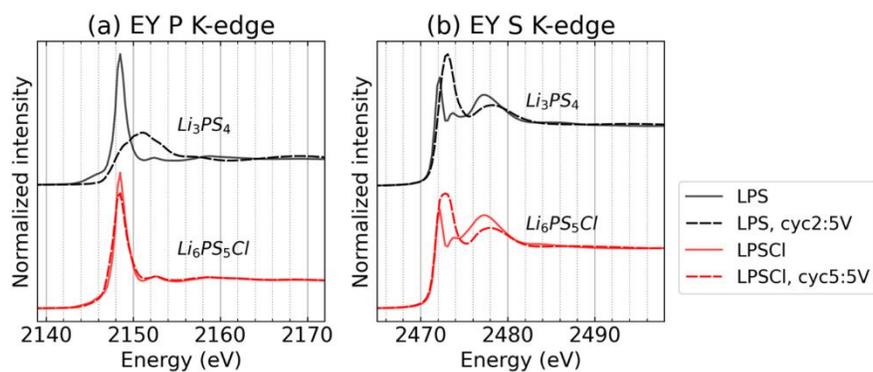

*Figure S14. (a) P K-edge and (b) S K-edge XANES data of sulfide electrolytes before and after cycling. In each panel, the black and red curves are from Li$_3$PS$_4$ and the argyrodite structured electrolyte Li$_6$PS$_5$Cl, respectively. The solid lines represent the pristine samples, and the dashed lines represent cycled samples. The data were collected in electron yield mode at SST-2 beamline at NSLS-II. The pristine LPS and LPS cyc2-5V data are also shown in Figure 2 (a) and (b) in the main text. Error in the measured XANES spectra is expected to be better than 0.1% and in the photon energy better than 0.05 eV.*

## 8. K-Means parameters and determining the number of clusters

We used the K-Means clustering method implemented in the Scikit-learn package[6]. The method for initialization is "k-means++" and the number of initializations is 10. The K-Means algorithm is "elkan" The max numter of iterations of the K-Means algorithm is 300 and the relative tolerance is 0.0001.

In the process of clustering the simulated spectra with K-Means, the number of clusters can in principle be any value. We use five clusters for phosphorus and six for sulfur, because these numbers give clustering results which can be reasonably intepreted, i.e., the structural characteristics in each cluster has unambiguous physical meaning. Furthermore, the numbers of clusters are corroborated by the elbow criterion[7].

Figure S15 are the Elbow plots for phosphorus and sulfur, showing the inertia versus the number of clusters, $k$. The reduction of inertia between adjcent $k$ values are also plotted. Inertia is defined as the sum of squared distances of samples to their closest cluster center[6, 8]. Three rounds of K-Means clustering are performed for each number of clusters to reduce the randomness. With small number of clusters, $k$, the inertia is large since the differences between individual spectrum and the cluster center are large. As $k$ increases, inertia decreases first rapidly, then the decrease slows down, forming an elbow shape. The turning point of this elbow shape can be used as an optimal $k$, which gives enough number of clusters for describing the data with small inner-cluster distances.

The turning point of sulfur is at $k$ = 6, which is in good agreement with our physics-based metrics. The turning point of phosphorus spectra is at $k$ = 7, whereas $k$ = 5 is used in our final results. Figure S16 plots the results using seven clusters. As a comparison, Cluster 2 and 3 in five-cluster results (Figure 4 in the main text) are further splited into Cluster 7-2, 7-3, 7-4, and 7-5 in seven-cluster results (Figure S16). The overall trend still maintains, i.e., larger distortion of $PS_4$ causes larger splitted white line peak. The optimal value of Elbow metrics is $k$ = 7, because the inner-cluster distances between spectra with differently splitted peaks are large. For simplicity, five clusters are used, because it provides the same information as seven clusters.

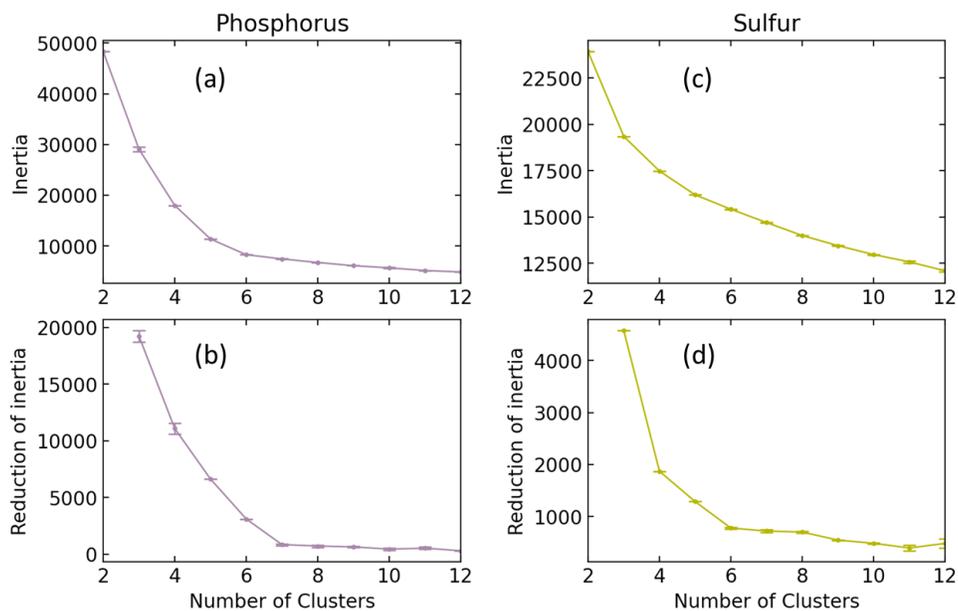

*Figure S15. Elbow plots for: (a, b) Phosphorus spectra, and (c, d) Sulfur spectra. (a) and (c) show the inertia versus cluster number k. (b) and (d) show the reduction of inertia versus, calculated as the reduction of inertia from the previous k value. The algorithm is run for three times at each k. For each run, the number of times the K-Means is run for different centroid seeds is set to 5 to compensate the randomness of choosing centroid seeds in K-Means algorithm. For each plot, the mean value is plotted with the solid line and the standard deviation from the three runs is plotted as the error bars.*

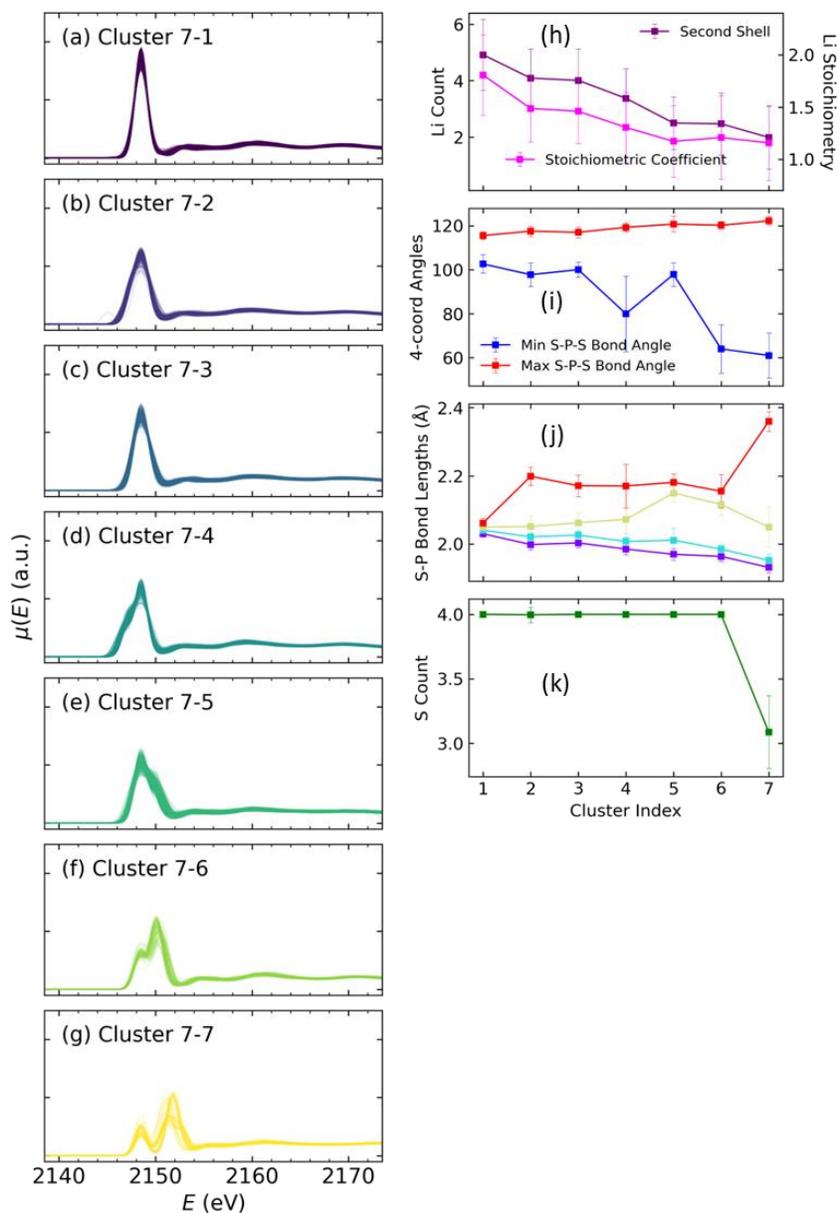

*Figure S16. Results using k = 7 in the clustering of phosphorus spectra. (a – g) Simulated P K-edge spectra grouped into seven clusters. (h – k) Averaged structural descriptors in each of the seven clusters, with markers showing the average values and error representing the standard deviation. We note that Cluster 5 has lower lithium stoichiometry than Cluster 6, but a higher number of Li atoms in the second shell. Since XAS measures the local structure motifs, whereas lithium stoichiometry is an averaged value over the entire supercell, we order the clusters by the second-shell lithium atoms.*